\documentclass[prodmode,acmtecs]{acmsmall} 

\usepackage[ruled]{algorithm2e}

\SetAlFnt{\small}
\SetAlCapFnt{\small}
\SetAlCapNameFnt{\small}
\SetAlCapHSkip{0pt}
\IncMargin{-\parindent}

\acmVolume{xx}
\acmNumber{xx}
\acmArticle{xx}
\acmYear{2013}
\acmMonth{10}

\usepackage[center,bf]{caption}
\usepackage[font=footnotesize]{subfig}

\usepackage{graphicx}
\graphicspath{{}}
\DeclareGraphicsExtensions{.eps,.pdf,.jpeg,.png}

\usepackage[cmex10]{amsmath}
\usepackage{url}
\usepackage{verbatim}

\usepackage{listings}
\newcommand{\code}[1]{\lstset{basicstyle=\small\tt}\lstinline£#1£\lstset
{basicstyle=\small\tt}}

\begin{document}
  
\markboth{I.Z. Reguly et al.}{Acceleration of a Full-scale Industrial CFD Application with OP2}

\title{Acceleration of a Full-scale Industrial CFD Application with OP2 \vspace{-0pt}}

\author{Istv\'an Z. Reguly
\affil{Oxford e-Research Centre, University of Oxford}
Gihan R. Mudalige
\affil{Oxford e-Research Centre, University of Oxford}
Carlo Bertolli
\affil{Department of Computing, Imperial College London}
Michael B. Giles
\affil{Oxford e-Research Centre, University of Oxford}
Adam Betts
\affil{Department of Computing, Imperial College London}
Paul H.J. Kelly
\affil{Department of Computing, Imperial College London}
David Radford
\affil{Rolls Royce plc.}\vspace{-0pt}
}

\begin{abstract}
Hydra is a full-scale industrial CFD application used for the design of turbomachinery at Rolls Royce plc. It consists
of over 300 parallel loops with a code base exceeding 50K lines and is capable of performing complex simulations over
highly detailed unstructured mesh geometries. Unlike simpler structured-mesh applications, which feature high speed-ups
when accelerated by modern processor architectures, such as multi-core and many-core processor systems, Hydra presents
major challenges in data organization and movement that need to be overcome for continued high performance on emerging
platforms. We present research in achieving this goal through the OP2 domain-specific high-level framework. OP2 targets
the domain of unstructured mesh problems and follows the design of an active library using source-to-source
translation and compilation to generate multiple parallel implementations from a single high-level application source
for execution on a range of back-end hardware platforms. We chart the conversion of Hydra from its original hand-tuned
production version to one that utilizes OP2, and map out the key difficulties encountered in the process. To our
knowledge this research presents the first application of such a high-level framework to a full scale production code.
Specifically we show (1) how different parallel implementations can be achieved with an active library framework, even
for a highly complicated industrial application such as Hydra, and (2) how different optimizations targeting contrasting
parallel architectures can be applied to the whole application, seamlessly, reducing developer effort and increasing
code longevity. Performance results demonstrate that not only the same runtime performance as that of the hand-tuned
original production code could be achieved, but it can be significantly improved on conventional processor
systems. Additionally, we achieve further acceleration by exploiting many-core parallelism, particularly on GPU systems.
Our results provide evidence of how high-level frameworks such as OP2 enable portability across a wide range of
contrasting platforms and their significant utility in achieving near-optimal performance without the intervention of
the application programmer. 

\end{abstract}

\category{C.4}{Performance of Systems}{Design Studies}
\terms{Design, Performance}
\keywords{
Unstructured Mesh Applications, Domain Specific Language, Active Library, OP2, OpenMP, GPU, CUDA, CFD \vspace{0pt}
}
\begin{bottomstuff}
Author's addresses: I.Z. Reguly,  G.R. Mudalige and M.B. Giles are with the Oxford e-Research Centre at the University
of Oxford, 7, Keble Road, Oxford OX1 3QG, UK. Email: \{istvan.reguly, gihan.mudalige\}@oerc.ox.ac.uk,
mike.giles@maths.ox.ac.uk. C. Bertolli (Current address) is with the IBM TJ Watson Research Centre, New York, USA.
Email:{cbertol@us.ibm.com}. A. Betts and P.H.J. Kelly are with the Dept. of Computing, Imperial College London UK.
Email:\{a.betts,p.kelly\}@imperial.ac.uk. D. Radford is with Rolls Royce plc. Derby UK.
Email:{David.Radford@rolls-royce.com}\vspace{5pt}
\end{bottomstuff}

\acmformat{Istv\'an Z. Reguly, Gihan R. Mudalige, Carlo Bertolli, Michael B. Giles, Adam Betts, Paul H. J. Kelly, and
David Radford, 2013. Acceleration of a Full-scale Industrial CFD Application with OP2.}

\maketitle

\vspace{-0pt}\section{Introduction}\label{sec/intro}

High Performance Computing (HPC) is currently experiencing a period of enormous change. For many years, increased
performance was achieved through higher clock frequencies giving an almost free boost to the speed and throughput of
applications without the need to re-write software for each new generation of processors. However, within the last
decade, further increase of clock frequencies and in turn higher performance was severely curtailed due to the rapid
increase in energy consumption. This phenomenon, commonly referred to as the end of Dennard's scaling~\cite{Bohr2007}, has
become significant as we reach the physical limits of the current CMOS based microprocessor technologies. The only clear
direction in gaining further performance improvements now appears to be through increased parallelism, where multiple
processor units are utilized to increase throughput. As a result, modern and emerging microprocessors feature multiple
homogeneous core optionally augmented with many simpler processing cores on a single piece of silicon; ranging from a few cores to
thousands, depending on the complexity of individual processing elements. For example, traditional mainstream processors
from Intel and AMD are continuing to be designed with more and more homogeneous cores. Each of these cores,
usually operating at a high clock frequency, in turn consists of increasingly large vector units (e.g. Intel's AVX
extensions) that simultaneously carry out operations on larger blocks of data. At the same time, there is an
emergence of co-processors designed to act as ``accelerators'' for augmenting the performance of certain types of
computations in combination with traditional processors. Examples of such co-processors range from  discrete (e.g. NVIDIA GPUs ~\cite{NVIDIA-GPUs}, Intel MIC~\cite{IntelMIC}) or integrated (e.g.
AMD APUs~\cite{AMDAPU}) SIMD type co-processors to more specialized DSP type compute engines~\cite{TI-DSP} or FPGA based
accelerators~\cite{FPGA-convey,FPGA-Maxceller}. The SIMD type accelerators currently consist of about 16-64 functional
units, each in turn consisting of a number of processor cores operating at a lower clock frequency that together can
perform a larger number of operations in parallel. Other HPC systems designs include the utilization of custom-built
large networks of relatively small but energy-efficient CPUs as in the IBM Blue Gene~\cite{BlueGeneQ} systems. In the
future, we may also see developments further motivated by energy-efficient designs from companies such as
ARM~\cite{ARM2011} who have achieved significant energy efficiency for mobile and embedded applications, and are now
targeting HPC which increasingly shares similar energy goals.

In light of these developments, an application developer faces a difficult problem. Optimizing an application for a
target platform requires more and more low-level hardware-specific knowledge and the resulting code is increasingly
difficult to maintain. Additionally, adapting to new hardware may require a major re-write, since optimisations
and languages vary widely between different platforms. At the same time, there is considerable uncertainty about which
platform to target: it is not clear which approach is likely to ``win'' in the long term. Application developers would
like to benefit from the performance gains promised by these new systems, but are concerned about the software
development costs involved.  Furthermore, it can not be expected of domain scientists to gain platform-specific
expertise in all the hardware they wish to use. This is especially the case with industrial applications that were
developed many years ago, often incurring enormous costs not only for development and maintenance but also for
validating and maintaining accurate scientific outputs from the application. Since these codes usually consist of
tens or hundreds of thousands of lines of code, frequently porting them to new hardware is infeasible. 

Recently, \emph{Active libraries}~\cite{Czarnecki-activelibrary,Veldhuizen-activelibrary} and \emph{Domain Specific
Languages} (DSLs) have emerged as a pathway offering a solution to these problems. The key idea is to allow scientists
and engineers to develop applications by providing higher-level, abstract constructs that make use of domain specific
knowledge to describe the problem to be solved. At the same time, appropriate code generation and compiler support will
be provided to generate platform specific code, from the higher-level source, targeting different hardware with tailored
optimisations. Within such a setting a separate lower implementation level is created to provide opportunities for
parallel programming experts to apply radically aggressive and platform specific optimizations when implementing the
required solution on various hardware platforms. The correct abstraction will pave the way for easy maintenance of a higher-level
application source with near optimal performance for various platforms and make it possible to easily integrate support
for any future novel hardware.

Much research~\cite{AEcute,liszt,Paraiso,Ypnos,SBLOCK,EPFL2011} has been carried out on such high-level abstraction
methods targeting the development of scientific simulation software. However, there has been no conclusive evidence, so
far, for the applicability of active libraries or DSLs in developing full scale industrial production applications,
particularly demonstrating the viability of the high-level abstraction approach and its benefits in terms of
performance and developer productivity. Indeed, it has been the lack of such an exemplar that has made these high level
approaches confined to university research labs and not a mainstream HPC software development strategy. This paper
presents research addressing this open question by detailing our recent experiences in the development and acceleration
of such an industrial application with the OP2 (~\cite{CJ2011,JPDC2012}) active library framework.

We focus on the industrial application Hydra, used at Rolls Royce plc. for the simulation of turbomachinery components
of aircraft engines. Hydra is a highly complex and configurable CFD application, capable of accommodating different
simulations that can be applied to any mesh. With the initial development carried out over 15 years ago, Hydra has been continuously evolving ever since. It is written in Fortran 77, and it is parallelized to
utilize a cluster of single threaded CPUs using message passing. Simulations implemented in Hydra are typically applied
to large meshes, up to tens of millions of edges, with execution times ranging from a few minutes up to a few weeks. Hydra uses a design based on a domain specific abstraction for the solution of unstructured mesh problems; the
abstraction is simply achieved with an API and its implementation is through a classical software library called OPlus
(Oxford Parallel Library for Unstructured Solvers)~\cite{oplus_b}. OPlus implements the API calls targeting a cluster
of single threaded CPUs. OP2, the second iteration of OPlus, was designed to retain the same domain specific abstraction
but develops an active library framework with code generation to exploit parallelism on modern heterogeneous multi-core
and many-core architectures.

With OP2, a single application code written using its API can be transformed (through source-to-source translation
tools) into multiple parallel implementations which can then be linked against the appropriate parallel library (e.g.
OpenMP, CUDA, MPI, OpenCL etc.) enabling execution on different back-end hardware platforms. At the same time, the
generated code and the OP2 platform specific back-end libraries are highly optimized utilizing the best low-level
features of a target architecture to make an OP2 application achieve near-optimal performance including high
computational efficiency and minimized memory traffic. In previous works, we have presented OP2's design and development
\cite{CJ2011,JPDC2012} and its performance on heterogeneous systems~\cite{InPar2012}. These works investigated the
performance through a standard unstructured mesh finite volume computational fluid dynamics (CFD) benchmark, called
``Airfoil'', written in C using the OP2 API and parallelized on a range of multi-core and many-core platforms; our
results showed considerable performance gains could be achieved on a diverse set of hardware.




In this paper we chart the conversion of Hydra from its original version, based on OPlus, to one that utilizes OP2, and
present key development and optimisation strategies that allowed us to gain near-optimal performance on modern parallel
systems. A key goal of this research is to investigate whether high-level frameworks such as OP2 could be used to
develop large-scale industrial applications and at the same time achieve performance on par with a hand-tuned
implementation. Specifically, we make the following contributions:
\vspace{-10pt}
\begin{enumerate}
\item \textit{Deployment}: We present the conversion of Hydra to utilize OP2, mapping out the key difficulties
encountered in the conversion of the Hydra application (designed and developed over 15 years ago) to OP2. Our
work demonstrates the clear advantages in developing future-proof and performant applications through a
\textit{high-level} abstractions approach. Hydra, to our knowledge, is the first industrial application to
successfully demonstrate the viability of such high-level frameworks.

\item \textit{Optimisations}: We present key optimisations that incrementally allowed Hydra to gain near optimal
performance on modern parallel systems, including conventional multi-core processors, many-core accelerators such as
GPUs as well their heterogeneous combinations. The optimisations are radically different across the range of platforms
under study, but through OP2, we are able to easily apply each new optimisation demonstrating portability and increased
developer productivity. 

\item \textit{Performance}: A range of performance metrics are collected to explore the performance of Hydra
with OP2, including runtime, scalability and achieved bandwidth. The performance is compared to that of the original
version of Hydra, contrasting the key optimisations that lead to performance differences. Benchmarked systems include a
large-scale distributed memory Cray XE6 system, and a distributed memory Tesla K20 GPU cluster interconnected by QDR
InfiniBand. The OP2 design choices and optimisations are explored with quantitative insights into their contributions
to performance on these systems. Additionally, performance bottlenecks of the application are isolated by breaking
down the runtime and analyzing the factors constraining total performance. 
\end{enumerate}

We use highly-optimized code generated through OP2 for all system back-ends, using the same application code, allowing
for direct performance comparison. Our work demonstrates how Hydra, through OP2, is developed to match the performance
of the original and then further outperform it with platform specific optimisations for modern multi-core
and accelerator systems. Re-enforcing our previous findings, this research demonstrates that an application written once
at a high-level using the OP2 framework is easily portable across a wide range of contrasting platforms, and is capable
of achieving near-optimal performance without the intervention of the application programmer.

The rest of this paper is organized as follows: Section~\ref{sec/background} introduces the Hydra CFD application;
Section~\ref{sec/deployment} present the transformations made to Hydra enabling it to utilize OP2;
Section~\ref{sec/perf} details performance and optimisations charting the effort to reach near optimal performance for
Hydra. Section~\ref{sec/hybrid} presents preliminary results from a hybrid CPU-GPU execution scheme.
Section~\ref{sec/relatedwork} briefly compares related work and Section~\ref{sec/conclusions} concludes the
paper. 

\vspace{-0pt}\section{Hydra}\label{sec/background}

The aerodynamic performance of turbomachinery is a critical factor in engine efficiency of an
aircraft, and hence is an important target of computer simulations. Historically, CFD simulations for turbomachinery
design were based on structured meshes, often resulting in a difference between simulation results and actual
experiments performed on engine prototypes. While the initial hypothesis for the cause was the poor quality of the
turbulence model, the use of unstructured meshes showed that the ability to model complex physical geometries
with highly detailed mesh topologies is essential in achieving correct results. As a result, unstructured mesh based
solutions are now used heavily to achieve accurate predictions from such simulations. 

Significant computational resources are required for the simulation of these highly detailed three-dimensional meshes. Usually the
solution involves iterating over millions of elements (such as mesh edges and/or nodes) to
reach the desired accuracy or resolution. Furthermore, unlike structured meshes, which utilize a regular stencil,
unstructured mesh based solutions use the explicit connectivity between elements during computation. This leads to very
irregular patterns of data access over the mesh, usually in the form of indirect array accesses. These data access
patterns are particularly difficult to parallelize due to data dependencies resulting in race conditions. 

Rolls Royce's Hydra CFD application is such a full-scale industrial application developed for the simulation of
turbomachinery. It consists of several components to simulate various aspects of the design including steady
and unsteady flows that occur around adjacent rows of rotating and stationary blades in the engine, the operation of compressors, turbines and exhausts as well as the simulation of behavior such as the
ingestion of ground vortexes. The guiding equations which are solved are the Reynolds-Averaged Navier-Stokes (RANS)
equations, which are second-order PDEs. By default, Hydra uses a 5-step Runge-Kutta method for time-marching,
accelerated by multigrid and block-Jacobi preconditioning~\cite{hydra4,hydra3,hydra2,LL2008}. The usual production meshes are in 3D and
consist of tens of millions of edges, resulting in long execution times on modern CPU clusters. 

Hydra was originally designed and developed over 15 years ago at the University of Oxford and has been in continuous development since, it has become one of the main
production codes at Rolls Royce. Hydra's design is based on a domain
specific abstraction for unstructured mesh based computations~\cite{oplus_b}, where the solution algorithm is separated into four
distinct parts: (1) sets, (2) data on sets, (3) connectivity (or mapping) between the sets and (4) operations over sets.
This leads to an API through which any mesh or graph problem solution can be completely and abstractly defined.
Depending on the application, a set can consist of nodes, edges, triangular faces, quadrilateral faces, or other
elements. Associated with these sets are data (e.g. node coordinates, edge weights, velocities) and mappings between
sets defining how elements of one set connect with the elements of another set. All the numerically intensive
computations can be described as operations over sets. Within an application code, this corresponds to loops over a
given set, accessing data through the mappings (i.e.~one level of indirection), performing some calculations, then
writing back (possibly through the mappings) to the data arrays.

The above API was implemented with the creation of a classical software library called OPlus, which provided a concrete
distributed memory parallel implementation targeting clusters of single threaded CPUs. OPlus essentially carried out
the distribution of the execution set of the mesh across MPI processes, and was responsible for all data movement while also taking
care of the parallel execution without violating data dependencies. As an example, consider the code in Figure 
\ref{fig/OPloop} of a loop in Hydra over a set of triangular cells as illustrated in Figure \ref{fig/examplemesh}:

\begin{figure}
\vspace{-0pt}\noindent\line(1,0){390}
\begin{lstlisting}
do while(op_par_loop(ncells, istart, iend))
  call op_access_r8('r',areac,1,ncells,
  &				null,0,0,1,1)
  call op_access_r8('u',arean,1,nnodes,
  &				ncell,1,1,1,3)
  do ic = istart, iend
	i1 = ncell(1,ic)
	i2 = ncell(2,ic)
	i3 = ncell(3,ic)
	arean(i1) = arean(i1) + areac(ic)/3.0
	arean(i2) = arean(i2) + areac(ic)/3.0
	arean(i3) = arean(i3) + areac(ic)/3.0
  end do
end while
\end{lstlisting}
\vspace{-5pt}\noindent\line(1,0){390}
\caption{\small A Hydra loop written using the OPlus API \normalsize}\vspace{-5pt}\label{fig/OPloop}
\end{figure}

The \texttt{op\_par\_loop} API call returns the loop bounds of the execution set, in this case triangular cells, while
the calls to \texttt{op\_access\_r8} update the halos at the partition boundary by carrying out MPI communications. The
actual numerical computation consists of distributing the area of each cell, \texttt{areac}, to the three nodes that
make up the cell. This is achieved by looping over each cell and accessing the nodes making up each cell indirectly
through the mapping \texttt{ncell} which points from the cell to its three nodes. The distributed memory parallel
implementation of the above loop is carried out by partitioning and distributing the global execution set on to
each MPI process. A single threaded CPU, assigned with one MPI process, will be sequentially iterating over its
execution set, from \textit{istart} to \textit{iend} to complete the computation. However, data dependencies at the
boundaries of the mesh partitions need to be handled to obtain the correct results. OPlus handles this by
creating suitable halo elements such that contributions from neighboring MPI processes are received to update the halo
elements. The implementation follows the standard MPI mesh partitioning and halo creation approach commonly found in distributed
memory MPI parallelizations. 

As it can be seen from the above loop, Hydra based on OPlus is tailored for execution on distributed memory single
threaded CPU clusters. However, as CPUs are increasingly designed with multiple processor cores and each with increasingly large
vector units, simply assigning an MPI process per core may not be a good long-term strategy if the full capabilities of
the processors are to be used. Moreover, based on experiences in attempting to exploit
the parallelism in emerging SIMD-type architectures such the Xeon Phi~\cite{IntelMIC} or GPUs, relying only on
coarse-grained message-passing is not going to be a viable or scalable strategy. Thus at least a thread-level, shared
memory based parallelism is essential if Hydra is to continue to perform well on future systems. On the other hand
further performance improvements appear to be obtainable with the use of accelerator based systems such as
clusters of GPUs. 



Directly porting the Hydra code to a multi-threaded implementation (e.g. with OpenMP or pthreads) is not
straightforward. Consider parallelizing the above loop with OpenMP. Simply adding a \texttt{!omp pragma parallel} for
the loop from \texttt{istart} to \texttt{iend} will not give correct results, due to the data races introduced by the
indirect data accesses through the \texttt{ncell} mapping. If higher performance is required more optimisations
tailored to OpenMP are needed. Thus, further substantial \textit{implementation-specific} modifications will be required
to achieve good thread-level parallelism. A much more significant re-write would be required if we were to get this loop
running on a GPU, say using CUDA. Again the application code base would be changed with significant implementation
specific code making it very hard to maintain. Porting to any future parallel architectures in this manner would yet
again involve significant software development costs. Considering that the full Hydra source consists of over 300 loops
written in Fortran 77, ``hand-porting'' in the above manner is not a viable strategy for each new type of parallel
system. The design of the OP2 framework was motivated to address this issue. 

\section{OP2}\label{sec/deployment}

While the initial motivation was to enable Hydra to exploit multi-core and many-core parallelism, OP2 was designed
from the outset to be a general high-level \emph{active library} framework to express and parallelize unstructured mesh
based numerical computations. OP2 retains the OPlus abstraction but provides a more complete high-level API (embedded in
C/C++ and Fortran) to the application programmer which for code development appears as an API of a classical software
library. However, OP2 uses a source-to-source translation layer to transform the application level
source to different parallelizations targeting a range of parallel hardware. This stage provides the opportunity to
provide the necessary implementation specific optimisations. The code generated for one of the
platform-specific parallelizations can be compiled using standard C/C++/Fortran compilers to generate the platform
specific binary executable. 

\subsection{The OP2 API}

\begin{figure}[t]\centering
\includegraphics[width=7cm]{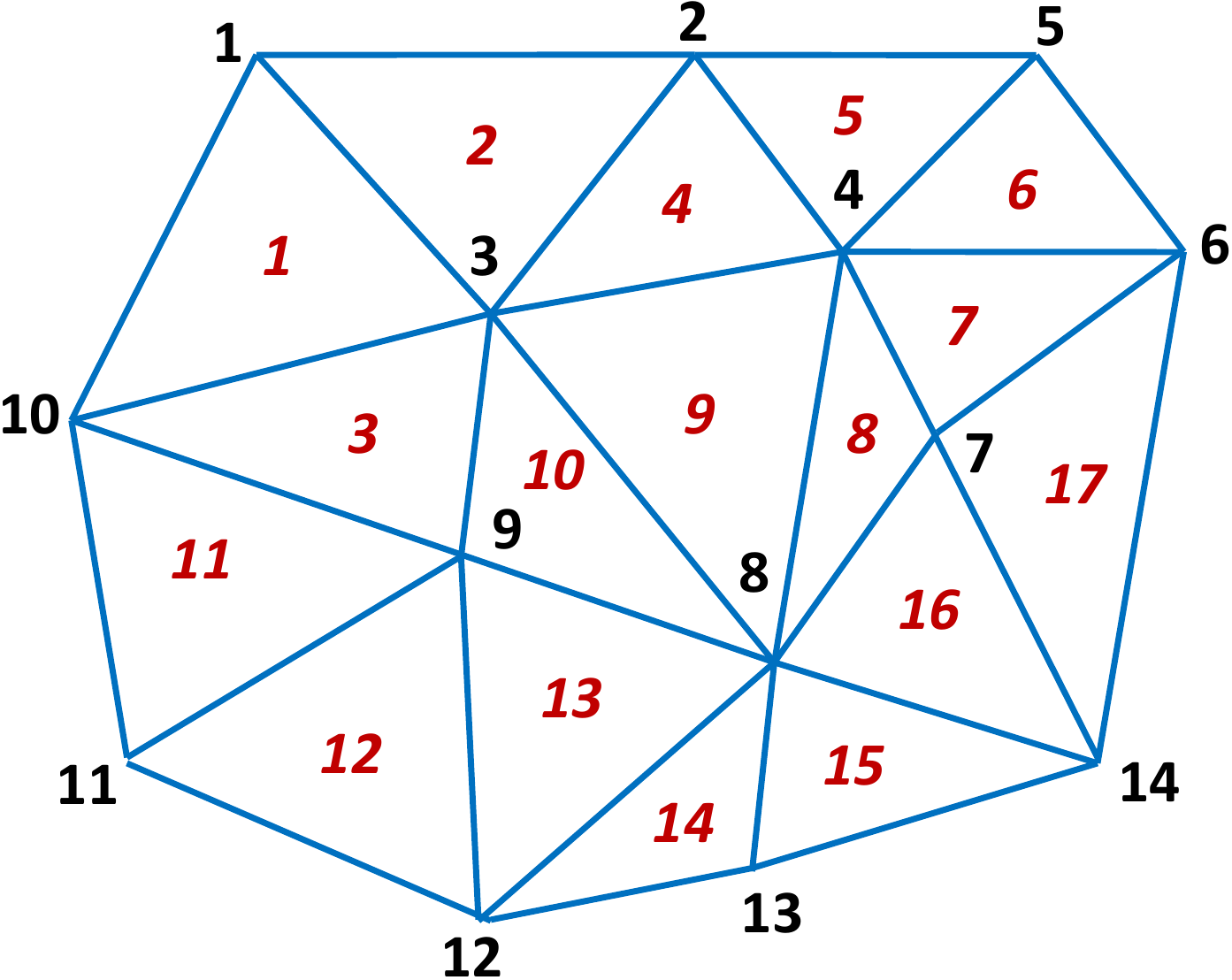}\vspace{-0pt}
\caption{\small An example unstructured mesh \normalsize}\vspace{-10pt}\label{fig/examplemesh}
\end{figure}

The first step in getting Hydra to exploit multi-core and many-core parallelism through OP2 is to convert it to use the
OP2 API. Continuing from the previous OPlus example, consider the unstructured mesh illustrated in Figure 
\ref{fig/examplemesh}. The mesh consists of three sets - nodes (vertices), edges and triangular cells. There are
\texttt{nn = 14} nodes and \texttt{nc = 17} cells. The OP2 API allows to declare these sets and connectivity
between the sets together with any data held on the sets (see Figure \ref{fig/declOP2}).

In \texttt{op\_decl\_map} each element belonging to the set \texttt{nodes} is mapped to three different elements in the
set \texttt{cells}. The \texttt{op\_map} declaration defines this mapping where \texttt{ncell} has a dimension of 3 and
thus the 1D array index 1,2 and 3 maps to nodes 1,3 and 10, index 4,5 and 6 maps to nodes 1,2 and 3 and so on. When
declaring a mapping we first pass the source set (e.g. \texttt{cells}) then the destination set (e.g. \texttt{nodes}).
Then we pass the dimension (or arity) of each map entry (e.g. 3; as \texttt{ncell} maps each cell to 3 nodes). Note that
those literal numbers filling up the mapping array are purely for description purposes only as OP2 supports input
from disk (e.g. using HDF5 files). Once the sets are defined, data can be associated with the sets through
\texttt{op\_decl\_dat} statements.~Note that here a single double precision value per set element is declared. A vector
of a number of values per set element could also be declared (e.g.~a vector with three doubles per node to store
coordinates). In the case of Hydra, there are \texttt{op\_dat}s with over 6 values per set element. 

\begin{figure}
\vspace{-0pt}\noindent\line(1,0){390}\vspace{-5pt}
\begin{lstlisting}
integer(4),dimension(:),allocatable:: c_to_n
real(8),dimension(:),allocatable:: ca_data
real(8),dimension(:),allocatable:: na_data

allocate ( c_to_n ( 3 * nc ) )
allocate ( ca_data ( nc ) )
allocate ( na_data ( nn ) )

c_to_n  = /1,3,10, 1,2,3, 3,9,10, 2,3,4, .../
ca_data = /.../
na_data = /.../

call op_decl_set(nn,nodes,'nodes')
call op_decl_set(ne,edges,'edges')
call op_decl_set(nc,cells,'cells')

call op_decl_map
&				(cells,nodes,3,c_to_n,ncell,'pcell')

call op_decl_dat
&				(cells,1,'r8',ca_data,areac,'c_area')
call op_decl_dat
&				(nodes,1,'r8',na_data,arean,'n_area')
\end{lstlisting}
\vspace{-10pt}\noindent\line(1,0){390}\vspace{0pt}
\caption{\small Declaring sets, maps and dats using the OP2 API \normalsize}\vspace{-5pt}\label{fig/declOP2}
\end{figure}

Any loop over the sets in the mesh is expressed through the \texttt{op\_par\_loop}, API call similar in purpose to the 
OPlus API call in Figure \ref{fig/OPloop} for a loop over \texttt{cells}. However, OP2 enforces a separation
of the per set element computation aiming to de-couple the declaration of the problem from the implementation. Thus,
the same loop can be expressed as detailed in Figure \ref{fig/OP2loop} using the OP2 API.

\begin{figure}
\vspace{-0pt}\noindent\line(1,0){390}\vspace{-5pt}
\begin{lstlisting}
subroutine distr(areac,arean1,arean2,arean3)
real(8), intent(in) :: areac
real(8), intent(inout) :: arean1,
	& arean2, arean3
arean1 = arean1 + areac/3.0
arean2 = arean2 + areac/3.0
arean3 = arean3 + areac/3.0
end subroutine

op_par_loop(cells, distr,
& op_arg_dat(areac,-1,OP_ID,1,'r8',OP_READ),
& op_arg_dat(arean,1,ncell,1,'r8',OP_INC),
& op_arg_dat(arean,2,ncell,1,'r8',OP_INC),
& op_arg_dat(arean,3,ncell,1,'r8',OP_INC))
\end{lstlisting}
\vspace{-10pt}\noindent\line(1,0){390}\vspace{0pt}
\caption{\small A Hydra loop written using the OP2 API \normalsize}\vspace{-5pt}\label{fig/OP2loop}
\end{figure}

The \texttt{subroutine distr()} is called a \textit{user kernel} in the OP2 vernacular. Simply put, the
\texttt{op\_par\_loop} describes the loop over the set \texttt{cells}, detailing the per set element computation as an
outlined ``kernel'' while making explicit indication as to how each argument to that kernel is accessed (OP\_READ - read
only, OP\_INC - increment) and the mapping, \texttt{ncell} (cells to nodes) with the specific indices are used to
indirectly accessing the data (\texttt{arean, areac}) held on each node. With this separation, the OP2 design, gives a
significantly larger degree of freedom in implementing the loop with different parallelization strategies. 

The one-off conversion of all the Hydra parallel loops to OP2's API simply involved extracting ``user-kernels'' from
each loop and then putting them in separate Fortran 90 modules. In this manner, the whole of Hydra was converted
consistently to use only the OP2 API. The conversion process was relatively straightforward due to the similarities of
the OPlus and OP2 APIs. Such a straightforward conversion may not have been possible if we were to convert a different
unstructured mesh application to use OP2. However, we believe that such a development cost is imperative for most
applications attempting to utilize the benefits of DSLs or Active Library frameworks. As we will show in this paper, the
advantages of such frameworks far outweigh the costs, by significantly improving the  maintainability of the application
source, while making it possible to also gain near optimal performance and performance portability across a wide range of
hardware.

\subsection{Code Generation and Parallel Build}

An application written with the OP2 API in the above manner can be immediately debugged and tested for accuracy by
including OP2's ``sequential'' header file (or its equivalent Fortran module if the application is written in Fortran).
This, together with OP2's sequential back-end library, implements the API calls for a single threaded CPU and can be
compiled and linked using conventional (platform specific) compilers (e.g. gcc, icc, ifort) and executed as a serial
application. OP2's CPU back-end libraries are implemented in C. To support applications developed with the Fortran API,
such as Hydra, the build process uses standard Fortran-to-C bindings, available since Fortran 2003. The Fortran
application code passes a Fortran procedure pointer and arguments to the \texttt{op\_par\_loop}. The module structure
for the sequential build is illustrated in Figure \ref{fig/OP2seqbuild}.

\begin{figure}
\vspace{-0pt}\noindent\line(1,0){390}\vspace{-5pt}
\begin{lstlisting}
! in file flux.F90
module FLUX
subroutine flux_user_kernel(x, ...)
real(8) x(3)
...
end subroutine
end module FLUX
\end{lstlisting}
\vspace{-10pt}\noindent\line(1,0){390}\vspace{0pt}
\vspace{-5pt}\noindent\line(1,0){390}\vspace{-0pt}
\begin{lstlisting}
! in file flux_app.F90
program flux_app
use OP2_Fortran_Reference
use OP2_CONSTANTS
use FLUX
...
call op_decl_set (nodes, ..)
call op_decl_map (..)
call op_decl_dat (x, ..)
...
call op_par_loop(nodes, flux_user_kernel,
  &  op_arg_dat(x,-1,OP_ID,3,'r8',OP_READ),
  &  ...)
...
end program flux_app
\end{lstlisting}
\vspace{-10pt}\noindent\line(1,0){390}\vspace{-0pt}
\caption{\small Modules structure for a sequential build with OP2 \normalsize}\vspace{-5pt}\label{fig/OP2seqbuild}
\end{figure}

In this illustration, the application consists of an \texttt{op\_par\_loop} that calls a Fortran 90 module called
\texttt{FLUX}. The module is in a separate file (\texttt{flux.F90}) and consists of the user kernel as a subroutine
called \texttt{flux\_user\_kernel}. The Fortran application code passes the \texttt{flux\_user\_kernel} procedure
pointer and arguments to the \texttt{op\_par\_loop}. The sequential implementation of the \texttt{op\_par\_loop} is
provided in the OP2 back-end library in the \texttt{OP2\_Fortran\_Reference} module. The calls to \texttt{op\_decl\_set,
op\_decl\_map} and \texttt{op\_decl\_dat} give OP2 full ownership of mappings and the data. OP2 holds them internally as
C arrays and it is able to apply optimizing transformations in how the data is held in memory. Transformations include
reordering mesh elements ~\cite{Burgess}, partitioning (under MPI) and conversion to an array-of-structs data layout (for
GPUs~\cite{JPDC2012}). These transformations, and OP2's ability to seamlessly apply them internally is key to achieving a
number of performance optimisations. 

\begin{figure}[t]\centering
\includegraphics[width=11cm]{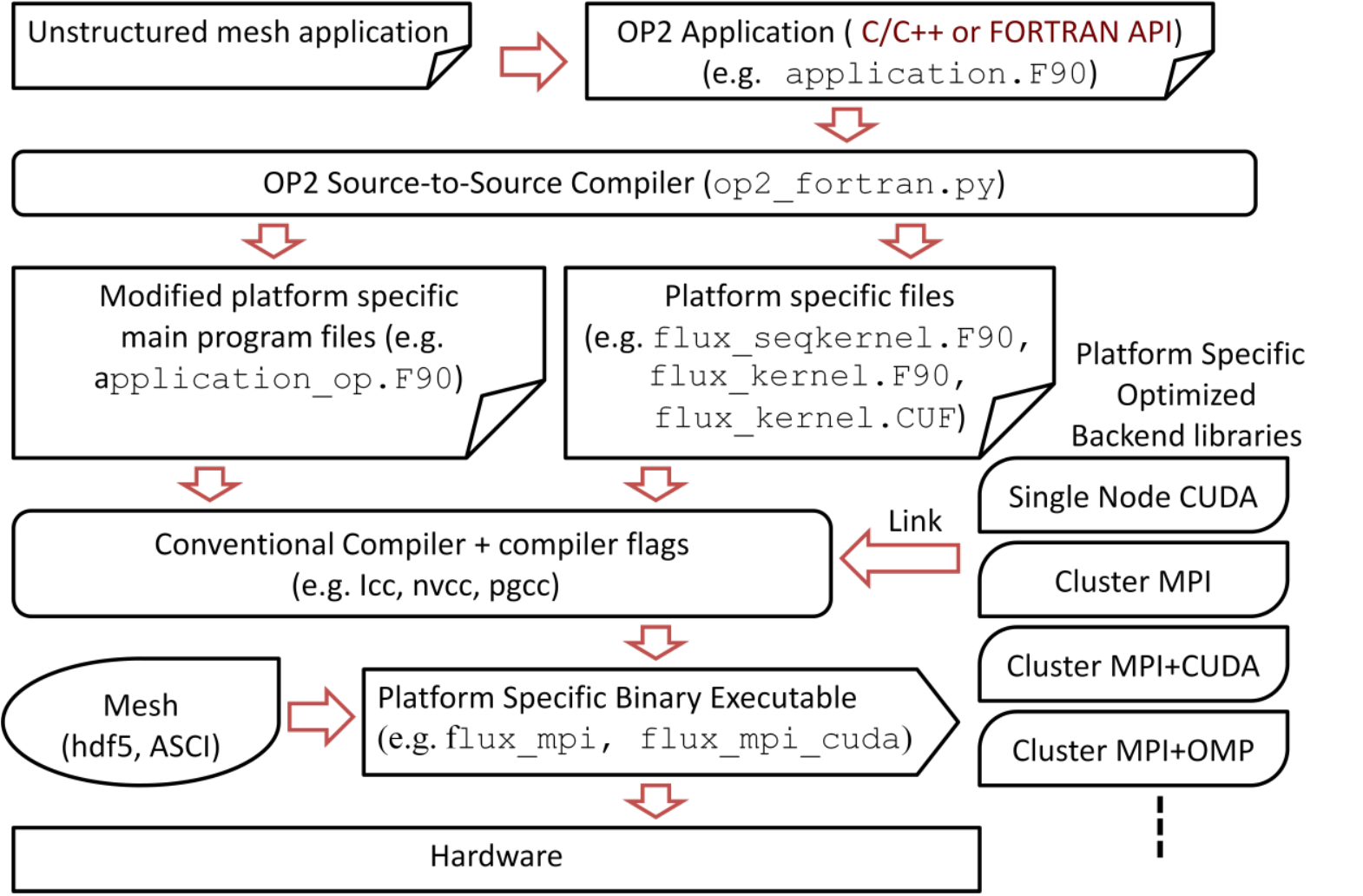}\vspace{-0pt}
\caption{\small OP2 build hierarchy \normalsize}\vspace{-5pt}\label{fig/parallelbuild}
\end{figure}

Once the application developer is satisfied with the validity of the results produced by the sequential application,
parallel code can be generated. The build process to obtain a parallel executable is detailed
in Figure \ref{fig/parallelbuild}. In this case the API calls in the application are parsed by the OP2
source-to-source translator which will produce a modified main program and back-end specific code. These are then
compiled using a conventional compiler (e.g. gcc, icc, nvcc) and linked against platform specific OP2 back-end libraries
to generate the final executable. The mesh data to be solved is input at runtime. The source-to-source code translator
is written in Python and only needs to recognize OP2 API calls; it does not need to parse the rest of the code.

OP2 currently supports parallel code generation to be executed on (1) multi-threaded CPUs/SMPs using OpenMP, (2) single
NVIDIA GPUs, (3) distributed memory clusters of single threaded CPUs using MPI, (4) cluster of multi-threaded CPUs using
MPI and OpenMP and (5) cluster of GPUs using MPI and CUDA. Race conditions that occur during shared-memory execution on
both CPUs (OpenMP) and GPUs (CUDA) are handled through multiple levels of coloring while for the distributed memory
(MPI) parallelization, an owner-compute model~\cite{PARCO2013} similar to that used in OPlus is implemented. More
details on the various parallelization strategies and their performance implications can be found in previous
papers~\cite{CJ2011,JPDC2012,InPar2012}.

For each parallelization, the generated modified main program and back-end specific code follow the general module
and procedure structure given in Figure \ref{fig/OP2parbuild}.
\begin{figure}
\vspace{-0pt}\noindent\line(1,0){390}\vspace{-5pt}
\begin{lstlisting}
!in file flux_seqkernel.F90 or flux_kernel.F90 or flux_kernel.CUF
module FLUX_HOST
use FLUX
subroutine flux_host_kernel(arg1, ...)
real(8), dimension(:), pointer :: arg1Ptr
call c_f_pointer(arg1%CPtr, arg1Ptr, ...)
...
! platform specific parallelisation 
do i = 0,nelems-1,1
  call flux_user_kernel(arg1Ptr(i*3+1,i*3+3),
  &  ...)
...
end subroutine
end module FLUX_HOST
\end{lstlisting}
\vspace{-10pt}\noindent\line(1,0){390}\vspace{0pt}
\vspace{-0pt}\noindent\line(1,0){390}\vspace{-5pt}
\begin{lstlisting}
!in file flux_app_op.F90
program flux_app_op
use OP2_FORTRAN_DECLARATIONS
use OP2_Fortran_RT_Support
use OP2_CONSTANTS
use FLUX_HOST
...
call flux_host_kernel(nodes, 
  & op_arg_dat(x,-1,OP_ID,3,'r8',OP_READ), 
  &  ...)
...
end program flux_app_op
\end{lstlisting}
\vspace{-10pt}\noindent\line(1,0){390}\vspace{-0pt}
\caption{\small Modules structure for a parallel build with OP2 \normalsize}\vspace{-5pt}\label{fig/OP2parbuild}
\end{figure}
A modified application program is generated in a new file called \texttt{flux\_app\_op.F90} which now uses the
OP2 back-end specific libraries implemented in \texttt{OP2\_FORTRAN\_DECLARATIONS} and
\texttt{OP2\_Fortran\_RT\_Support} modules. In this case the \texttt{op\_par\_loop} API call is converted to a
subroutine named \texttt{flux\_host\_kernel}, which is implemented in a module called \texttt{FLUX\_HOST}. The parallel
implementation of the \texttt{op\_par\_loop} in the \texttt{flux\_host\_kernel} subroutine differs for each
parallelization (e.g MPI, OpenMP, CUDA etc.). As such the subroutine \texttt{flux\_host\_kernel} is placed in a separate
file \texttt{flux\_seqkernel.F90}, \texttt{flux\_kernel.F90} or \texttt{flux\_kernel.CUF} depending on whether we are
generating a single threaded CPU cluster parallelization (with MPI), multi-threaded CPU cluster parallelization (with
OpenMP and MPI) or cluster of GPUs parallelization (with CUDA and MPI). The \texttt{flux\_host\_kernel} subroutine in
turn calls the user kernel \texttt{flux\_user\_kernel}. In the code illustration in Figure \ref{fig/OP2parbuild}
we have shown the MPI only parallelization. The calls to \texttt{c\_f\_pointer} are necessary to bind the C pointers,
which point to the C data and mapping arrays held internally by OP2, to Fortran pointers in order to pass them to the
user kernel. This allows OP2 to use the user kernels without modification by the code generator.

One final step that was required to get the Fortran API working with the OP2 back-end was the handling of global
constants. The original Hydra code with OPlus used common blocks to declare and hold global constants where their value
is set during an initialization phase and then used throughout the code. With the move to use F90 in OP2, all
constants were declared in a separate module \texttt{OP2\_CONSTANTS}. For the GPU implementation, in order to separate
constants held in the device (i.e. GPU) and the host CPU the \texttt{constant} keyword in the variable type qualifiers
(alternatively \texttt{device} if the array is too large) and the \texttt{\_OP2CONSTANT} suffix was added to the names of constant
variable. In Hydra, all global variables that use common blocks are defined in a number of header files,
thus we were able to implement a simple parser that extracts variable names and types, and generates the required
constants module.


At this stage, the conversion of Hydra to utilize the OP2 framework was complete. The application was validated to check
that the correct scientific results were obtained. The open question now then was whether the time and effort spent in the
conversion of Hydra to utilize an active library such as OP2 is justified. Specifically the key questions were: (1)
whether the conversion to OP2 gave Hydra any performance improvements compared to the original OPlus version, (2)
whether further performance gains are achievable with modern multi-core/many-core hardware (3) what optimisations can be
applied to improve performance for different parallel platforms and (4) how the OP2 framework facilitate the deployment
of such optimisations. In the next section we analyze Hydra's performance with OP2 and present work assessing these
issues.


\section{Performance and Optimisations}\label{sec/perf}

We begin by initially benchmarking the runtime of Hydra with OP2 on a single node. Exploring performance on a single
node is not only motivated by the need to understand intra-node performance but also due to the fact that Hydra is
regularly run on single node systems by CFD engineers for preliminary design purposes. Key specifications of the single
node system are detailed in column 1 of Table \ref{tab/systems}. The system is a two socket Intel Xeon E5-2640
system with 64GB of main memory. The processors are based on Intel's latest Sandy Bridge architecture. The compiler
flags that give the best runtimes are listed. This system, named Ruby, also consists of two NVIDIA Tesla K20c GPUs, each
with 5 GB of graphics memory. We use CUDA 5.0 in this study. Table \ref{tab/systems} also details the
large cluster systems used later in the benchmarking study. These will be used to explore the distributed memory scaling
performance of Hydra.

\begin{table*}[t]\small
\centering
\caption{Benchmark systems specifications \normalsize}
\begin{tabular}{cccc} \hline
System	    &Ruby 				       &HECToR 			    &Jade       	      \\
	    &(Development machine)		       &(Cray XE6)		    &(NVIDIA GPU Cluster)     \\\hline
Node	    &$2\times$Tesla K20c GPUs+		       &$2\times$16-core AMD Opteron&$2\times$Tesla K20m GPUs+    \\
Architecture&$2\times$6-core Intel Xeon &6276 (Interlagos)2.3GHz	    &Intel Xeon \\
& E5-2640 2.50GHz & & E5-1650 3.2GHz\\\hline
Memory/Node &5GB/GPU + 64GB		       & 32GB		            &5GB/GPU        \\\hline
Num of Nodes&1		       		      	       & 128		            & 8        		     \\\hline
Interconnect& shared memory			       & Cray Gemini	 	    &FDR InfiniBand          \\\hline
O/S	    &Red Hat Linux 6.3	       & CLE 3.1.29		    &Red Hat Linux 6.3	
\\\hline
Compilers   &PGI 13.3, ICC 13.0.1,		       &  Cray MPI 8.1.4            & PGI 13.3, ICC 13.0.1,  \\
 	    &OpenMPI 1.6.4, CUDA 5.0			       &			    & OpenMPI 1.6.4, CUDA 5.0     \\\hline
Compiler    &-O2 -xAVX			  	       & -O3 -h fp3 -h ipa5 	    & -O2 -xAVX   	     \\
flags 	    &-Mcuda=5.0,cc35	       &  	 		    &-Mcuda=5.0,cc35
\\\hline
\end{tabular}\label{tab/systems}
\end{table*}\normalsize

\begin{figure}[t]\centering
 \includegraphics[width=7.5cm]{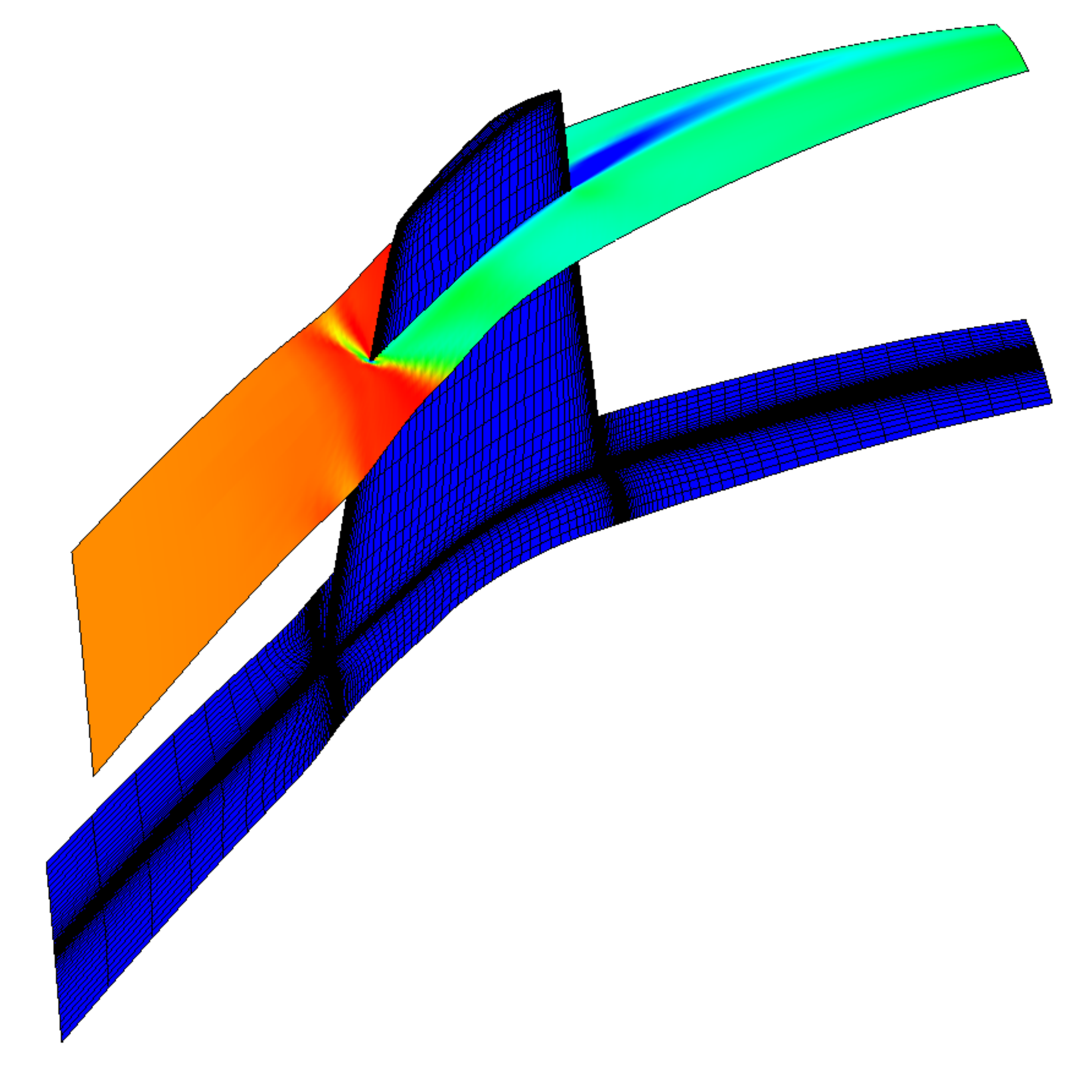}
 \vspace{-5pt}\caption{\small Mach contours for NASA Rotor 37.\normalsize}
 \label{fig:nasa}\vspace{-15pt}
 \end{figure}

As mentioned previously, Hydra consists of several components~\cite{LL2008} and in this paper we report on the
non-linear solver configured to compute in double precision floating point arithmetic. Hydra can also be used with
multi-grid simulations, but for simplicity of the performance analysis and reporting we utilize experiments with a
single grid (mesh) level. The configuration and input meshes of Hydra in these experiments model a standard application
in CFD, called NASA Rotor37~\cite{nasarotor37}. It is a transonic axial compressor rotor widely used for validation in
CFD. Figure \ref{fig:nasa} shows a representation of the Mach contours for this application on a single blade passage.
The mesh used for the single node performance benchmarking consists of 2.5 million edges.

\begin{figure*}[!t]
\begin{center}
\subfloat[OPlus vs OP2 (MPI only)\label{fig/rubyperf}]{\includegraphics[width=8.5cm]{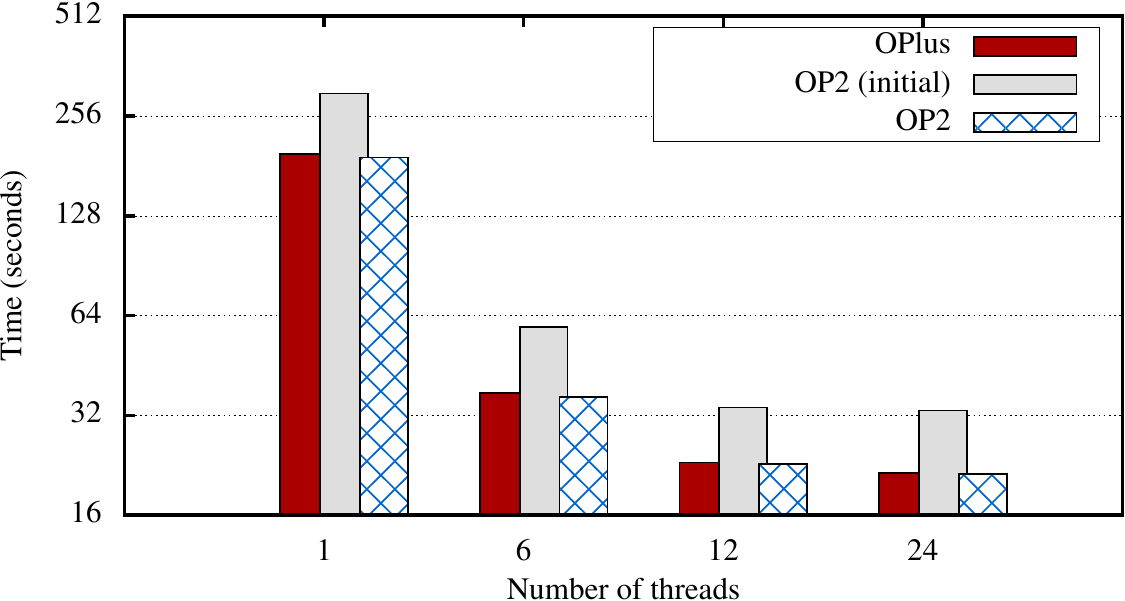}}\\\vspace{-10pt}
\subfloat[OPlus vs OP2 (with PTSotch and renumbering)\label{fig/rubyperf2}]
{\includegraphics[width=8.5cm]{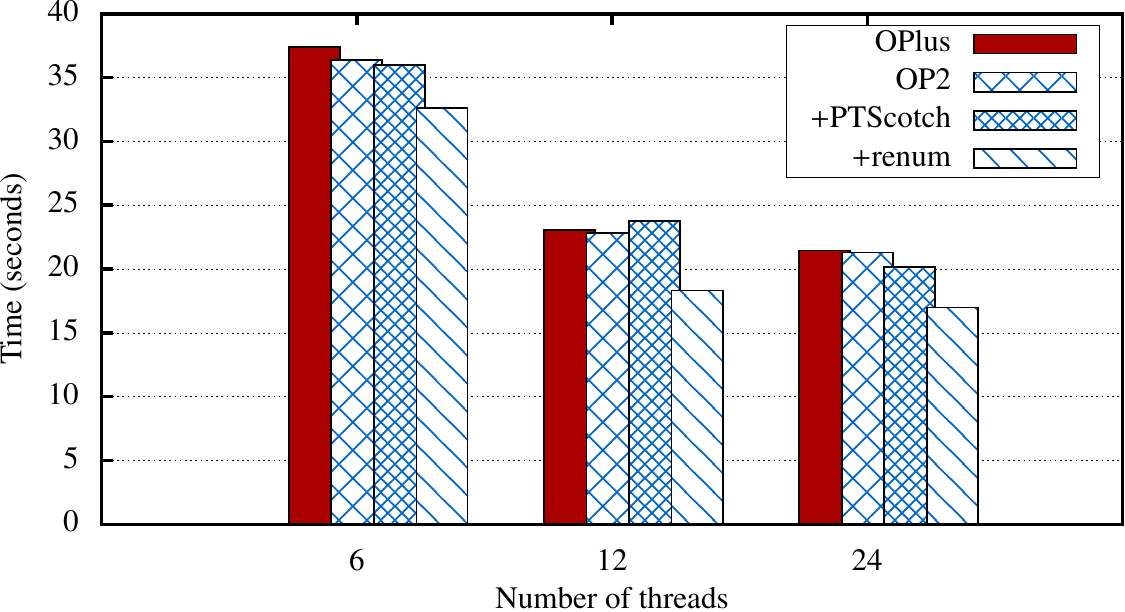}}\vspace{-5pt}
\caption{Single node performance on Ruby (NASA Rotor 37, 2.5M edges, 20 iterations)}%
\end{center}\vspace{-15pt}
\end{figure*}

Figure \ref{fig/rubyperf} presents the performance of Hydra with both OPlus and OP2 on up to 12 cores (and 24
SMT threads) on the Ruby single node system using the message passing (MPI) parallelization. This is a like-for-like
comparison where the same mesh is used by both versions. The partitioning routine used in both cases is a recursive
coordinate bisection (RCB) mesh partitioning~\cite{rcb} where the 3D coordinates of the mesh are repeatedly split in the
x, y and z directions respectively until the required number of partitions (where one partition is assigned to one MPI
process) is achieved. The timings presented are for the end-to-end runtime of the main time-marching loop for 20
iterations. Usual production runs solving this mesh would take hundreds of iterations to converge. 

We see that the OP2 version (noted as OP2 initial) is about 50\% slower than the hand coded OPlus version. The generated
code from OP2 appears to be either missing a performance optimisation inherent in the original Hydra code and/or the
OP2 generated code and build structure is introducing new bottlenecks. By simply considering the runtime on a single
thread we see that even without MPI communications the OP2 (initial) version performs with the same slowdown. Thus it
was apparent that some issue was affecting single threaded CPU performance. One plausible explanation was that the
generated files and the separation of the user kernel is inhibiting function in-lining optimizations. As a solution, the
code generator was modified to generate code that simply places the user kernel and the encompassing loop over set
elements into the same source file, adding compiler pragmas to ensure the inlining of the user kernel. The resulting
code indeed improved performance over the OP2 (initial) version on average by about 25\%. 

\begin{figure}
\vspace{-0pt}\noindent\line(1,0){390}\vspace{-5pt}
\begin{lstlisting}
module FLUX_GENSEQ
subroutine flux_user_kernel(x, ...)
real(8) x(3)
...
end subroutine

subroutine flux_wrap(arg1data,...,nelems)
real(8), arg1data(3,*)
do i = 0,nelems-1,1
  call flux_user_kernel(x(1,i+1),
  &  ...)
enddo
end subroutine

subroutine flux_kernel(arg1, ...)
real(8), dimension(:), pointer :: arg1Ptr
...
call c_f_pointer(arg1%CPtr, arg1Ptr, ...)
...
call flux_wrap(arg1Ptr,...)
...
end subroutine
end module FLUX_GENSEQ
\end{lstlisting}
\vspace{-10pt}\noindent\line(1,0){390}\vspace{0pt}
\vspace{-0pt}\noindent\line(1,0){390}\vspace{-0pt}
\begin{lstlisting}
!in file flux_app_op.F90
program flux_app_op
use OP2_FORTRAN_DECLARATIONS
use OP2_Fortran_RT_Support
use OP2_CONSTANTS
use FLUX_GENSEQ
...
call flux_kernel(nodes, 
  & op_arg_dat(x,-1,OP_ID,3,'r8',OP_READ),
  &  ...)
...
end program flux_app_op
\end{lstlisting}
\vspace{-10pt}\noindent\line(1,0){390}\vspace{0pt}
\caption{\small Optimized modules structure for OP2 
\normalsize}\vspace{-15pt}\label{fig/OP2optbuild}
\end{figure}


To analyze the performance further, we ran Hydra through the Intel VTune~\cite{vtune} profiler. The aim was to
investigate the performance of each subroutine that is called when executing an \texttt{op\_par\_loop}. As Hydra
consists of more than 300 parallel loops, we focused on one of the most time consuming \texttt{op\_par\_loop}s called
\texttt{edgecon}. The profiling revealed that a significant overhead (up to 40\% of the total instruction count for the
loop) occurs at the call to the user kernel; for example, the call to the \texttt{flux\_user\_kernel} subroutine in
Figure \ref{fig/OP2parbuild}. Further investigation revealed that the cause is a low level Fortran specific
implementation issue~\cite{ISOF_to_C} related to how arrays are represented internally in Fortran. The pointer 
\texttt{arg1Ptr} in Figure \ref{fig/OP2parbuild}, results in an ``assumed shape'' array which is represented
by an internal struct called a ``dope vector'' in Fortran. The dope vector contains the starting pointer of the array
but also bounds and stride information. The extra information facilitates strided array sections and features such as
bounds checking. However computing the memory address of an array element held in the struct causes a complex indexing
calculation. Thus, directly using this array pointer and passing it as an argument to the \texttt{flux\_user\_kernel}
subroutine causes the indexing calculation to occur for each and every iteration in the \texttt{do i = 0,nelems-1,1}
loop.

The resolution is to force the complex index calculation to occur only once. After some experimentation, a modified
module and subroutine structure as illustrated in Figure \ref{fig/OP2optbuild} was generated. The new structure
introduces a wrapper subroutine \texttt{flux\_wrap} that is initially called by \texttt{flux\_kernel} by passing the
\texttt{arg1Ptr} pointer. At this point the complex indexing calculation is carried out, just once. Then
\texttt{flux\_wrap} works with an ``assumed shape array'' and calls the user kernel \texttt{flux\_user\_kernel} with a
simple offset that is cheap to calculate. To apply the modification to all parallel loops we simply modified the code
generator to generate code with this new subroutine structure. None of the application level code with the OP2 API was
affected. The performance of the resulting code is presented by the third bar in Figure~\ref{fig/rubyperf}. The
OP2 based Hydra application now matches the performance of the original version. These results by themselves
provide enormous evidence of the utility of OP2, particularly (1) showing how an optimization can be applied to the
whole industrial application seamlessly through code generation, reducing developer effort and (2) demonstrating that
the same runtime performance as that of a hand-tuned code could be achieved through the high-level framework.


A number of new features implemented in OP2 allow further improvements for the MPI parallelization. Firstly,
the OP2 design allows the underlying mesh partitioner for distributing the mesh across MPI processes to be changed. During
the initial input, OP2 distributes the sets, maps and data assuming a trivial block partitioning, where consecutive
blocks of set elements are assigned to consecutive partitions. As this trivial partitioning is not optimized
for distributed memory performance, a further re-partitioning is carried out (in parallel) with the use of
state-of-the-art unstructured mesh partitioners such as ParMetis~\cite{parmetis} or PTScotch~\cite{ptscotch}. For the
purposes of comparison with the original OPlus code, as shown in Figure \ref{fig/rubyperf}, we have also implemented the
recursive coordinate bisection algorithm in OP2 allowing it to generate the exact same partitioning as the original
Hydra application. 

Secondly, OP2 allows the ordering or numbering of mesh elements in an unstructured mesh to be optimized. The renumbering of
the execution set and related sets that are accessed through indirections has an important effect on performance~\cite{Burgess}: cache
locality can be improved by making sure that data accessed by elements which are executed consecutively are close, so
that data and cache lines are reused. OP2, implements a renumbering routine that can be called to convert the
input data meshes based on the Gibbs-Poole-Stockmeyer algorithm in Scotch~\cite{ptscotch}.


The results in Figure \ref{fig/rubyperf2} show the effect of the above two features. The use of PTScotch 
resulted in about 8\% improvement over the recursive coordinate bisection partitioning, on Ruby. However,
renumbering resulted in about 17\% gains for Hydra solving the NASA Rotor 37 mesh. Overall, partitioning with PTScotch
gave marginally better performance than ParMetis (not shown here). In the results presented in the rest
of the paper we make use of OP2's mesh renumbering capability, unless stated otherwise.

Breakdowns for some of the most time-consuming loops are shown in Table \ref{tab/loopperfmpi} when all the above optimisations are applied; observe how only 6 loops make up 75\% of the total runtime. The table also shows the data requirements per set element; the number of double precision numbers read and written, either directly or indirectly, ignoring temporary memory requirements due to local variables. The results suggest that the memory system is under considerable pressure, especially when large amounts of data is accessed indirectly, but also shows that performance depends on the computations carried out within the user kernel. Further benchmarks exploring different aspects of performance are presented in Section~\ref{sec/perfscale} on large-scale systems. 

 \begin{table}[t]\small
 \centering\vspace{-5pt}
 \caption{Hydra single node performance with 24 MPI processes, showing data requirements per set element as number of double precision values read and written either directly or indirectly (NASA Rotor 37, 0.8M nodes, 2.5M edges, 20 iterations).}\vspace{0pt}
 \begin{tabular}{lrrrr}
 Loop	   		& Time (sec)& \% runtime & Direct R/W & Indirect R/W\\\hline
 \texttt{accumedges}	&1.32	&7.83 & 3/0 &74/52\\
 \texttt{edgecon}	&1.08	&6.56 & 3/0 & 68/48\\ 
 \texttt{ifluxedge}	&1.49	&8.96 & 3/0 & 34/12\\
 \texttt{invjacs}	&0.17	&1.03 & 27/27 & 0/0\\
 \texttt{srck}		&0.35	&1.86 & 30/6 & 0/0\\
 \texttt{srcsa}		&1.32	&7.60 & 34/6 & 0/0\\
 \texttt{updatek}	&0.98	&5.83 & 53/6 & 0/0\\
 \texttt{vfluxedge}	&6.52	&38.58 & 3/0 & 92/12\\
 \texttt{volap}		&0.43	&2.58 &29/24 &0/0\\
 \texttt{wffluxedge}	&0.23	&1.38 & 7/0 & 72/12\\
 \texttt{wvfluxwedge}	&0.21	&1.24 & 4/0 & 45/6\\\hline
 \end{tabular}
 \label{tab/loopperfmpi}\vspace{-0pt}
 \end{table}\normalsize



\subsection{Multi-core and Many-core Performance}\label{subsec/perfmulti}

OP2 supports parallel code generation for execution on multi-threaded CPUs or SMPs using OpenMP and on NVIDIA GPUs
using CUDA. The generated OpenMP code uses the same subroutine, module and file structure illustrated in
Figure \ref{fig/OP2optbuild}. At the time of writing, CUDA and NVIDIA's compiler, nvcc only supports code
development in C/C++, so OP2 utilizes PGI's CUDA FORTRAN compiler to support code generated for the GPU via the Fortran
API.

To exploit shared memory parallelization techniques, OP2 segments the execution set on a partition into blocks
or mini-partitions and each mini-partition is assigned to an OpenMP thread (or a CUDA thread block) for execution in
parallel~\cite{CJ2011,InPar2012}. However, to avoid race conditions due to indirectly accessed data the blocks are
colored such that adjacent blocks are given different colors. When executing the computations per block, only blocks of
the same color are executed in parallel; furthermore, on the GPU, a subsequent coloring of set elements within each
block is necessary to avoid race conditions when one set element is assigned to one GPU thread. All these form an
execution plan that is created for each loop when it is first encountered and then reused during subsequent executions.

\subsubsection{OpenMP Performance}\label{subsubsec/openmp}

\begin{figure}[t]\centering
\includegraphics[width=8.4cm]{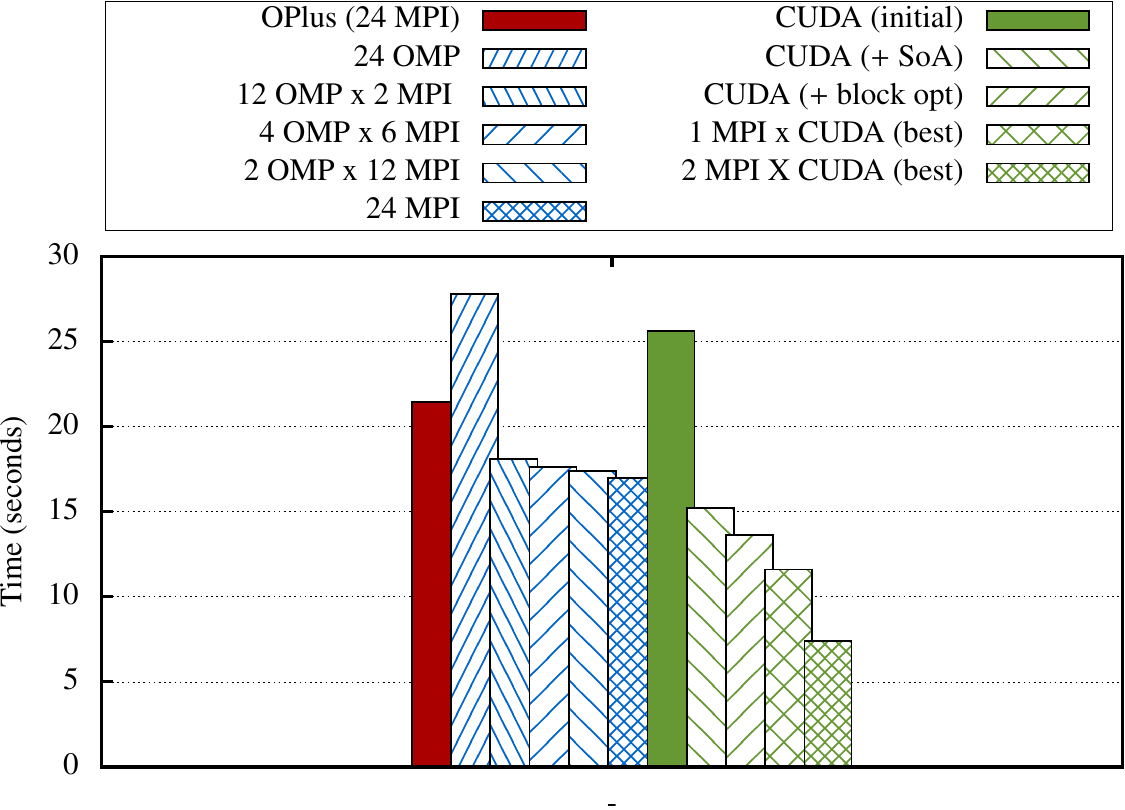}
\vspace{-5pt}\caption{OP2 Hydra Multi-core/Many-core performance on Ruby (NASA Rotor 37, 2.5M edges, 20 iterations) with
PTScotch partitioning}
\label{fig/RubyCUDA}\vspace{-10pt}
\end{figure}

Figure \ref{fig/RubyCUDA} presents the runtime performance of the OpenMP parallel back-end. The experiments varied from
running a fully multi-threaded version of the application (OpenMP only), to a heterogeneous version using both MPI and
OpenMP. 

\begin{figure}[t]\centering
\includegraphics[width=8.4cm]{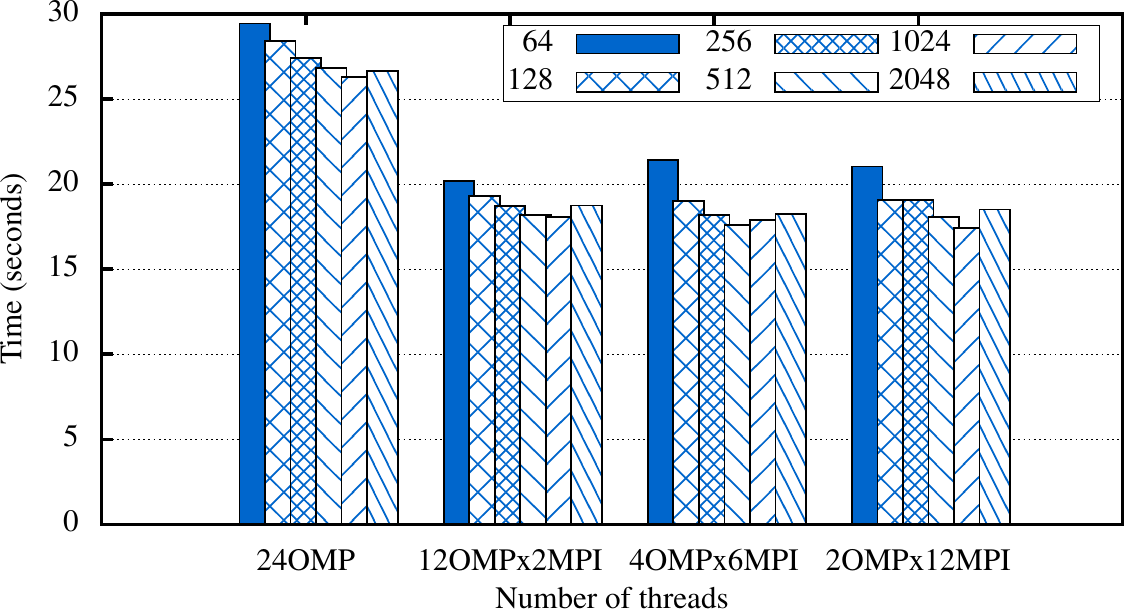}
\vspace{-5pt}\caption{Hydra OpenMP and MPI+OpenMP performance for different block sizes on Ruby (NASA Rotor 37, 2.5M
edges, 20 iterations) with PTScotch partitioning}
\label{fig/OMP-blocksizes}\vspace{-5pt}
\end{figure}

We see that executing Hydra with 24 OpenMP threads (i.e. OpenMP only) resulted in significantly poorer performance than
when using only the MPI parallelization, on this two socket CPU node. We have observed similar performance with the
Airfoil CFD benchmark code~\cite{PARCO2013}. The causes for MPI outperforming OpenMP on SMPs/CMPs have been widely
discussed in literature. These mainly consist of the non-uniform memory access issues (NUMA~\cite{Lameter2013,NUMA})
discussed in further detail below, and thread creation and tear down overheads~\cite{threading}. In our case, an
additional cause may also be the reduced parallelism due to coloring for blocks that avoid data races. The hybrid
MPI+OpenMP parallelization provided better performance but was still about 10\% slower than the pure MPI version; again
the overhead of shared-memory multi-threading techniques may be causing performance bottlenecks. To explore the causes
further, in Figure~\ref{fig/OMP-blocksizes} and Table~\ref{tab/OMP-blks} we present the variation in the runtime and
the number of block colors for different block sizes for a number of combinations of MPI processes and OpenMP threads.
The number of colors and blocks per color are automatically output by OP2 for each \texttt{op\_par\_loop} and we have
presented here a range of numbers taken from the most time consuming loops in Hydra. 

With the shared memory parallelism execution scheme employed by OP2, we reason that the size of
the block determines the amount of work that a given thread carries out uninterrupted. The bigger it is, the higher data
reuse within the block with better cache and prefetch engine utilization. At the same time, some parallelism is lost
due to the colored execution. In other words, only those blocks that have the same color can be executed at the same
time by different threads, with an implicit synchronization step between colors. This makes the execution scheme prone
to load imbalances, especially when the number of blocks with a given color is comparable to the number of threads that
are available to execute them.

The above two causes have given rise to the runtime behavior we see from each MPI and OpenMP combination in
Figure \ref{fig/OMP-blocksizes}. The average number of colors (nc) and the average number of blocks (nb) in 
Table \ref{tab/OMP-blks} supports this reasoning where, overall we see a reduction in runtime as the block
size is increased but only until there are enough blocks per color (nb/nc) to achieve good load balancing in
parallel. We also experimented by using OpenMP's static and dynamic load balancing functionality but did not obtain any
significant benefits as all the blocks executed (except for the very last one) have the same size.

The overall poor performance of the pure OpenMP runtimes (24OMP) for any block size, compared to the other MPI+OpenMP
runs, appears to be due to Non-Uniform Memory Accesses (NUMA) issues. On multiple-socket machines system memory is
connected to different sockets, and so when a thread on one socket has to access memory connected to
another socket there is a significant overhead ~\cite{Lameter2013,NUMA}. As OP2 does a colored execution, and the
colors are dynamically determined at runtime and may vary between different loops, statically determining which thread
processes which blocks is not possible, thus allocating memory close to the thread/core that will execute a block is
not feasible. Thus, the pure OpenMP version gets affected, preventing further performance gains from increased block
size. When executing in an MPI+OpenMP hybrid setting, processes and threads are pinned to specific sockets, thereby
circumventing this issue. Thus it is important to use a sensible MPI and OpenMP process and thread combination for the
given node/socket architecture.  

Finally we conclude that the above effects, particularly the lost parallelism with colored execution, combined with
thread creation and tear down overheads~\cite{threading} may account for the slightly worse performance of the
hybrid MPI+OpenMP approach on a single node compared to the MPI only runtime. 

\begin{table}[t]\small
\centering
\begin{center}
\vspace{0pt} 
\end{center}
\begin{tabular}{l|rr|rr|rr|rr}
	&\multicolumn{8}{c}{OMP $\times$ MPI} \\\hline
Block  	&
\multicolumn{2}{c|}{24}&\multicolumn{2}{c|}{12$\times$2}&\multicolumn{2}{c|}{4$\times$6}&\multicolumn{2}{c}{2$\times$12
} \\
size		& nb	& nc	& nb	& nc	& nb	& nc	&  nb	& nc\\
		& (K)	& 	& (K)	& 	& (K)	& 	&  (K)	& \\\hline
64		& 39	& 14	& 20	& 16 	& 6.8	& 16	&  3	& 16\\\hline
128		& 20	& 15	& 10	& 17	& 3.4	& 17	&  1.7	& 15\\\hline
256		& 10	& 14	& 5	& 14	& 1.7	& 15	&  0.8	& 14\\\hline
512		& 5	& 12	& 2.5	& 14	& 0.8	& 14	&  0.4	& 12\\\hline
1024		& 2.5	& 10	& 1	& 11	& 0.4	& 11	&  0.2	& 11\\\hline
2048		& 1	& 9	& 0.6	& 11	& 0.2	& 11	&  0.1	& 10\\\hline
\end{tabular}
\caption{Hydra single node performance, Number of blocks (nb) and \\number of colors (nc) (K = 1000): 2.5M
edges, 20 iterations}
\label{tab/OMP-blks}\vspace{-5pt}
\end{table}\normalsize

Table \ref{tab/loopperfOMP} details the achieved bandwidth for the most time consuming loops of Hydra with OpenMP
(6 MPI $\times$ 4 OMP). These loops together take up about 90\% of the total runtime. The achieved bandwidths are
reported through code generated by OP2 for performance reporting. The bandwidth figure was computed by counting the
useful amount of data bytes transferred from/to global memory during the execution of a parallel loop and dividing it by
the runtime of the loop. The achieved bandwidth by a majority of the key loops on Ruby is over or close to 60\% of the
advertised peak bandwidth of the Sandy Bridge processors (2$\times$42.6 GB/s ~\cite{SandyBridge}). The best performing
loops, such as \texttt{srck, updatek} and \texttt{volapf}, are direct loops that only access datasets defined on the
set being iterated over, hence they achieve a very high fraction of the peak bandwidth, due to trivial access patterns
to memory. Indirect loops use mapping tables to access memory, and they often require colored execution in order to
avoid data races; both of these factors contribute to lower bandwidth achieved by loops such as \texttt{accumedges,
ifluxedge} and \texttt{edgecon}. Additionally some loops are also compute and control-intensive, such as
\texttt{vfluxedge}. Finally, loops over boundary sets, such as \texttt{wvfluxedge}, have highly irregular access
patterns to datasets and generally much smaller execution sizes, leading to a lower utilization of resources. Our
experiments also showed that the trends on achieved bandwidth utilization remain very similar to the observed results
from previous work~\cite{CJ2011,InPar2012} for the Airfoil benchmark application.

\begin{table}[t]\small
\centering\vspace{10pt}
\caption{Hydra single node performance, 6 MPI x 4 OMP\\ with PTScotch: 2.5M edges, 20
iterations}\vspace{0pt}
\begin{tabular}{lrrr}
Loop	   		& Time (sec)&  GB/sec & \% runtime \\\hline
\texttt{accumedges}	& 1.46	&28.57	&7.99 \\
\texttt{edgecon}	& 2.00	&58.40	&10.95 \\
\texttt{ifluxedge}	& 1.88	&48.88	&10.32 \\
\texttt{invjacs}	& 0.16	&67.37	&0.90 \\
\texttt{srck}		& 0.47	&81.72	&2.57 \\
\texttt{srcsanode}	& 1.34	&31.62	&7.38 \\
\texttt{updatek}	& 0.94	&68.67	&5.14 \\
\texttt{vfluxedge}	& 7.05	&16.11	&38.70 \\
\texttt{volapf}		& 0.47	&72.19	&2.57 \\
\texttt{wffluxedge}	& 0.26	& 25.24	&1.41 \\
\texttt{wvfluxedge}	& 0.26	& 16.81	&1.41 \\\hline
\end{tabular}
\label{tab/loopperfOMP}\vspace{-5pt}
\end{table}\normalsize

\subsubsection{GPU Performance}\label{subsubsec/gpu}

The Ruby development machine contains two Tesla K20 GPUs, using which we investigate Hydra's performance on GPUs. Many
aspects of the generated CUDA code had been optimized from our previous work~\cite{CJ2011,JPDC2012} targeting the older
generation of NVIDIA GPUs based on their Fermi architecture. Thus, we re-evaluated the generated CUDA code targeting the
Tesla K20 which are based on NVIDIA's latest Kepler architecture. 

In a way similar to the code generation for the CPUs, we use Fortran to C bindings to call functions in the OP2 back-end
and to connect C pointers to Fortran pointers. The same plan construction that was described in previous papers
\cite{JPDC2012} is retained. Previously, indirect data from GPU global memory were loaded into each SM’s shared memory
space forming a local mini-partition. Each mini-partition was assigned to an SM (Streaming multi-processor) and 
executed in parallel. The SM executes the mini-partition by utilizing a number of threads (called a thread block).
However, we observed that this staging of indirectly accessed data is not beneficial for Hydra contrary to the results
we observed with the Airfoil benchmark~\cite{JPDC2012} on previous-generation hardware. This is due to the fact that
the large amounts of data moved by parallel loops in Hydra would require excessive amounts of shared memory, which in turn
would severely degrade multiprocessor occupancy (the number of threads resident in a multiprocessor at the same time)
and performance. Therefore, we eliminated the staging in shared memory by directly loading the data from global memory
into SMs, and rely on the increased L2 and texture cache size to speed up memory accesses that have spatial and temporal
locality. The performance achieved by the CUDA application generated by OP2 with such a parallel implementation is
presented in Figure \ref{fig/RubyCUDA}. However, this initial code on a single GPU ran about 45\% slower than the best
CPU performance on Ruby (2 CPUs, 24 MPI processes). 


Running the generated CUDA code through the NVIDIA Visual Profiler~\cite{NSIGHT} revealed more opportunities for
optimizations. It appeared that the switch to directly loading data from global memory (without staging in shared
memory) has made most parallel loops in Hydra limited by bandwidth utilization. Further investigation showed that a high
amount of cache contention is caused by the default data layout of \texttt{op\_dat}s. This layout, called
Array-of-Structs (AoS), was found to be the best layout for indirectly accessed data in our previous
work~\cite{JPDC2012} on Fermi GPUs. However without staging in shared memory it was damaging performance in Hydra as
many of Hydra's \texttt{op\_dat}s have a large number of components (dimensions) for each set element (typically related
to the number of PDEs $\geq 6$). Thus, adjacent threads are accessing memory that are far apart, resulting in high
numbers of cache line loads (and evictions, since the cache size is limited). OP2 has the ability to effectively
transpose these datasets and use a Struct-of-Arrays (SoA) layout, so when adjacent threads are accessing the same data
components, they have a high probability of being accessed from the same cache line. This is facilitated by either the
user annotating the code or by telling the OP2 framework to automatically use the SoA layout for datasets above a given
dimension. The above optimisation resulted in an increase of about 40\% to the single GPU performance as shown in
Figure \ref{fig/RubyCUDA}. 


To improve performance further, two other aspects of GPU performance were investigated. The goal was to allow as many
threads to be active simultaneously so as to hide the latency of memory operations. The first consideration is to limit
the number of registers used per thread. A GPU's SM can hold at most 2048 threads at the same time, but it has only  a
fixed number of registers available (65k 32-bit registers on Kepler K20 GPUs). Therefore kernels using excessive amounts of
registers per thread decrease the number of threads resident on an SM, thereby reducing parallelism and performance.
Limiting register count (through compiler flags) can be beneficial to occupancy and performance if the spilled
registers can be contained in the L1 cache. Thus for the most time-consuming loops in Hydra, we have manually adjusted
register count limitations so as to improve occupancy. 

The second consideration is to adjust the number of threads per block. Thread block size not only affects occupancy,
but also has non-trivial effects on performance due to synchronization overheads, cache locality, etc. that are
difficult to predict. We used auto-tuning techniques described in a previous work \cite{JPDC2012} to find the best
thread block size for different parallel loops, and stored them in a look-up table. This optimisation can be carried out
once for different hardware, but performance is unlikely to vary significantly when solving different problems. 
Together with the SoA optimisation, the use of tuned block sizes and limited register counts provided a further 10\%
performance improvement. 

\begin{table}[t]\small
\centering\vspace{-5pt}
\caption{Hydra single GPU performance:\\NASA Rotor 37, 2.5M edges, 20 iterations}\vspace{0pt}
\begin{tabular}{lrrr}
Loop	   		& Time (sec)&  GB/sec & \% runtime \\\hline
\texttt{accumedges}	&2.07	&13.87	&19.30 \\
\texttt{edgecon}	&2.46	&33.55	&22.87 \\
\texttt{ifluxedge}	&1.03	&71.90	&9.59 \\
\texttt{invjacs}	&0.41	&25.94	&3.85 \\
\texttt{srck}		&0.34	&108.09	&3.20 \\
\texttt{srcsa}		&0.34	&121.72	&3.15 \\
\texttt{updatek}	&0.54	&115.91	&5.01 \\
\texttt{vfluxedge}	&2.32	&53.82	&21.62 \\
\texttt{volap}		&0.23	&142.85	&2.14 \\
\texttt{wffluxedge}	&0.11	&31.55	&1.00 \\
\texttt{wvfluxwedge}	&0.08	&25.69	&0.77 \\\hline
\end{tabular}
\label{tab/loopperfCUDA}\vspace{-5pt}
\end{table}\normalsize

In the case of several Hydra loops that are bandwidth-intensive and do little computation, the thread coloring technique
(that avoids write conflicts) introduces a significant branching overhead, due to adjacent edges (mapped to adjacent
threads) having different colors, and execution over colors getting serialized. The negative effect on performance can
be mitigated by sorting edges in blocks by color, ensuring that threads in a warp have the same color (or at least fewer
colors), which in turn reduces branching within the warps. However, this solution degrades the access patterns to
directly accessed data in memory, as adjacent threads are no longer processing adjacent set elements. Thus, we only
apply it to the Hydra loops that are limited by branching overhead. The final best runtime on a single K20 GPU and on both
the K20 GPUs on Ruby is presented in the final two bars of Figure \ref{fig/RubyCUDA}. A single K20 GPU achieves about
1.8$\times$ speedup over the original OPlus version of Hydra on Ruby and about 1.5$\times$ over the MPI version of Hydra
with OP2. However, considering the full capabilities of a node, the best  performance on Ruby with GPUs (2 GPUs with
MPI) is about 2.34$\times$ speedup over the best OP2 runtime with CPUs (2 CPUs, 24 MPI processes) and about 2.89$\times$
speedup over the best OPlus runtime with CPUs (2 CPUs, 24 MPI processes). Table \ref{tab/loopperfCUDA} notes the
achieved memory bandwidth utilization on a K20 GPU. A majority of the most time consuming loops achieve 20 - 50 \% of the
advertised peak bandwidth of 208 GB/s on the GPU~\cite{k20}. Bandwidth utilization is particularly significant during
the direct loops \texttt{srck, updatek} and \texttt{volapf}. 

As it can be seen from the above effort, a range of low-level features had to be taken into account and a significant
re-evaluation of the GPU optimizations had to be done to gain optimal performance even when going from one generation of
GPUs by the same vendor/designer to the next. In our case the previous optimizations implemented for the Fermi GPUs had to be
considerably modified to achieve good performance on the next generation Kepler GPUs. However, as the code was developed
under the OP2 framework, radical changes to the parallel code could be easily implemented. In contrast, a directly
hand-ported application would cause the application programmer significant difficulties to maintain performance for each
new generation of GPUs, not to mention new processor architectures. This shows that the use of OP2 has indeed led to a
near optimal GPU back-end for Hydra that is significantly faster than the CPU back-end, with very little change to the
original source code.

\subsection{Distributed Memory Scaling Performance}\label{sec/perfscale}

The industrial problems simulated by Hydra require significantly larger computational resources than what is available
today on single node systems. An example design simulation such as a multi-blade-row unsteady RANS (Reynolds Averaged
Navier Stokes) computation where the unsteadiness originates from the rotor blades moving relative to the stators,
would need to operate over a mesh with about 100 million nodes. Currently with OPlus, Hydra can take more than a week on a small CPU cluster to reach convergence for such a large-scale problem. Future turbomachinery design
projects aim to carry out such simulations more frequently, such as on a weekly or on a daily basis, and as such the OP2
based Hydra code needs to scale well on clusters with thousands to tens of thousands of processor cores. In this section we explore the performance on such systems. Table \ref{tab/systems} lists
the key specifications of the two cluster systems we use in our benchmarking. The first system, HECToR~\cite{hector}, is
a large-scale proprietary Cray XE6 system which we use to investigate the scalability of the MPI and MPI+OpenMP
parallelizations. The second system, JADE~\cite{jade} is a small NVIDIA GPU (Tesla K20) cluster that we use to benchmark
the MPI+CUDA execution.


\begin{figure*}[!t]
\begin{center}
\subfloat[Strong Scaling (2.5M edges)]{\includegraphics[width=8.5cm]{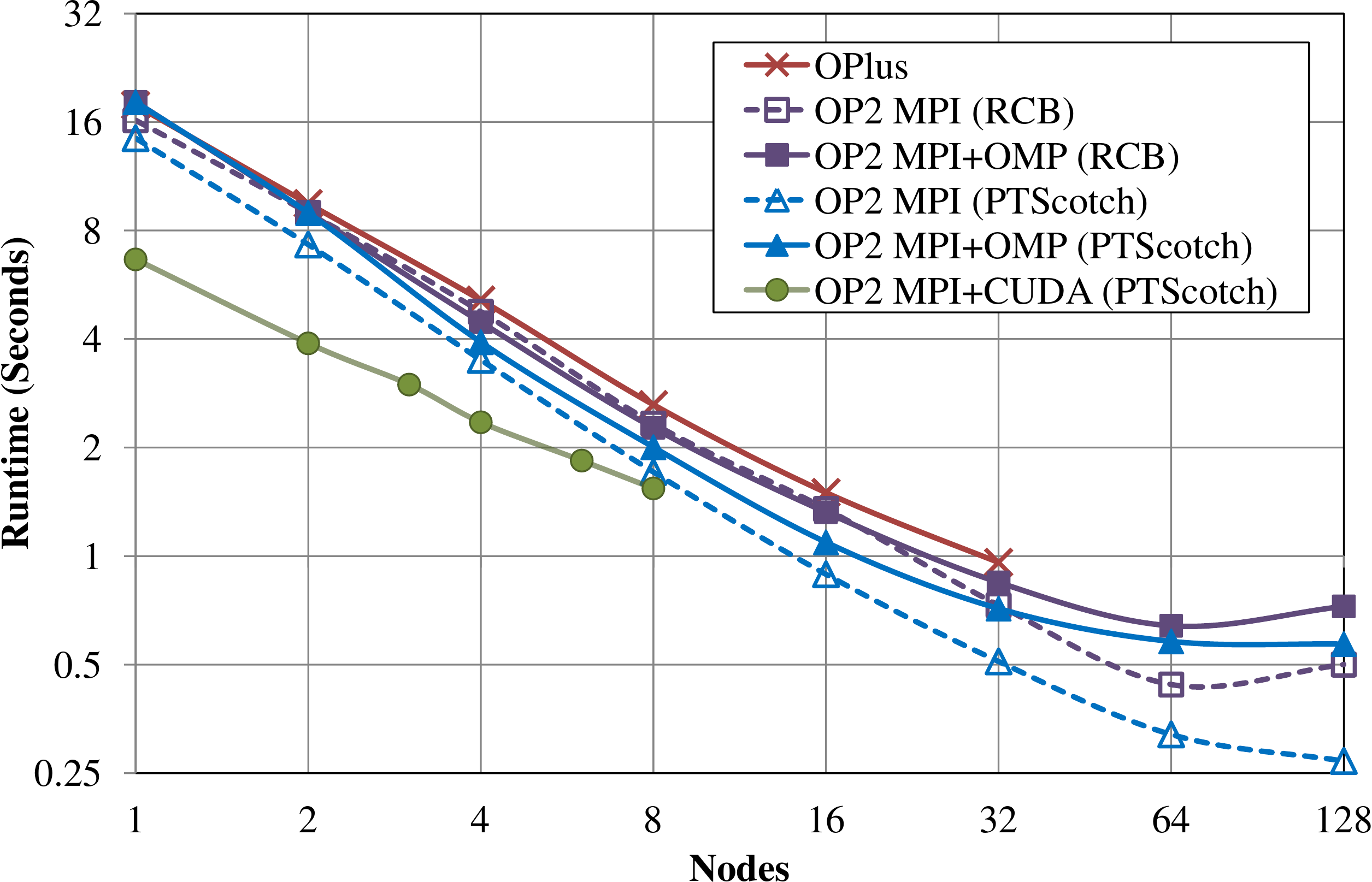}}
\\\vspace{-10pt}
\subfloat[Weak Scaling (0.5M edges per node)]{\includegraphics[width=8.5cm]{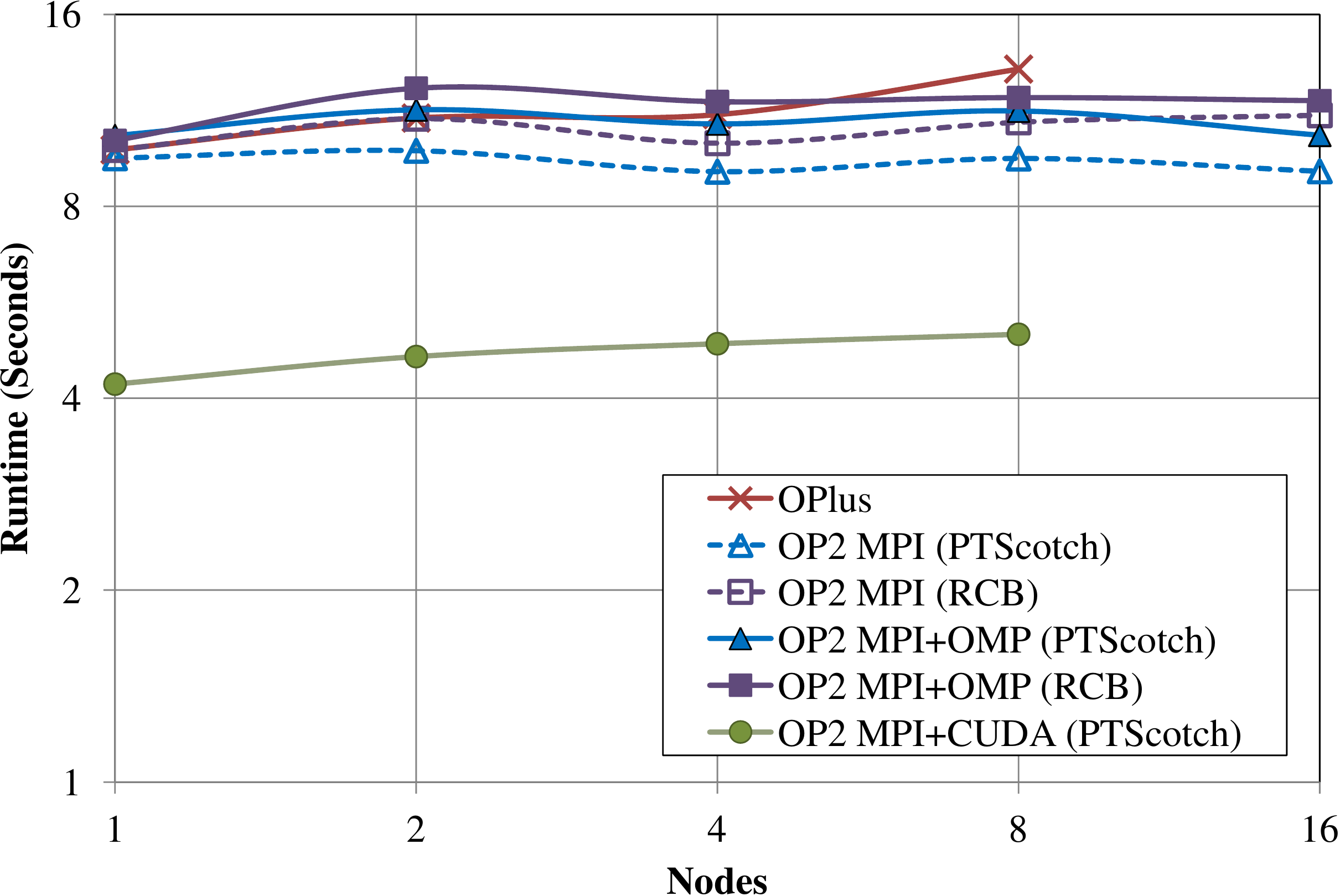}}\vspace{-0pt}
\caption{Scaling performance on HECToR (MPI, MPI+OpenMP) and Jade (MPI+CUDA) on the NASA Rotor 37 mesh (20
iterations)}%
\label{fig/scaling}
\end{center}\vspace{-15pt}
\end{figure*}

\begin{table*}[t]\small
\centering
\caption{Halo sizes : Av = average number of MPI neighbors per process,
\\Tot. = average number total elements per process, \%H = average \% of \\halo elements per
process}\label{tab/halos}
\subfloat[Strong Scaling]{
\centering
\begin{tabular}{rr|rrr|rrr}\hline
nodes 	&MPI    & \multicolumn{3}{c|}{PTScotch}	&\multicolumn{3}{c}{RCB}   \\ 
	&procs. & Av	& Tot 	& \%H 	&Av	& Tot 	& \%H    \\\hline
1	&32	&7	&111907	&5.12	&9	&117723	&9.81 \\
2	&64	&8	&56922	&6.74	&10	&61208	&13.27 \\
4	&128	&9	&29175	&9.02	&11	&32138	&17.41 \\
8	&256	&10	&14997	&11.52	&11	&17074	&22.28 \\
16	&512	&10	&7765	&14.55	&12	&9211	&27.97 \\
32	&1024	&10	&4061	&18.32	&12	&4949	&32.98 \\
64	&2048	&10	&2130	&22.16	&13	&2656	&37.58 \\
128	&4096	&10	&1134	&26.98	&13	&1425	&41.89 \\\hline
\end{tabular}
}\hspace{0pt}
\subfloat[Weak Scaling]{
\centering
\begin{tabular}{rr|rrr|rrr}\hline
nodes 	&MPI    & \multicolumn{3}{c|}{PTScotch}	&\multicolumn{3}{c}{RCB}   \\ 
	&procs. & Av	& Tot 	& \%H 	&Av	& Tot 	& \%H    \\\hline
1	&32	&8	&70084	&6.00	&8	&73486	&10.35 \\
2	&64	&9	&73527	&6.69	&9	&78934	&11.07 \\
4	&128	&10	&79009	&6.09	&10	&73794	&12.26 \\
8	&256	&12	&73936	&7.33	&11	&78396	&12.98 \\
16	&512	&13	&78671	&6.94	&12	&75224	&13.76 \\\hline\\\\\\
\end{tabular}}
\vspace{-25pt}
\end{table*}\normalsize

\subsubsection{Strong Scaling}\label{subsubsec/strongscaling}

Figure \ref{fig/scaling}(a) reports the run-times of Hydra at scale, solving the NASA Rotor 37 mesh with 2.5M edges
in a strong-scaling setting. The x-axis represents the number of nodes on each system tested, where a HECToR node consists
of two Interlagos processors, a JADE node consists of two Tesla K20 GPUs. The run-times (on the y-axis) are averaged
from 5 runs for each node count. The standard deviation in run times was always less than 10\%. MPI+OpenMP
results were obtained by assigning four MPI processes per HECToR node, each MPI process consisting of eight OpenMP
threads. This is due to the NUMA architecture of the Interlagos processors~\cite{interlagos_whitepaper} which combines
two cores to a ``module'' and packages 4 modules with a dedicated path to part of the DRAM. Here we were able to
leverage the job scheduler to exactly place and bind one MPI process per memory region (or die) reducing inter-die
communications. Other combinations of processes and threads were also explored, but the above provided the best
performance. 

On HECToR, we see that the overall scaling of Hydra with OP2 is significantly better than that with OPlus.
OP2's MPI-only parallelization scales well up to 128 nodes (4096 cores). At 32 nodes (1024 cores) the MPI-only
parallelization partitioned with PTScotch gives about 2x speedup over the runtime achieved with OPlus. 

As with all message passing based parallelizations, one of the main problems that limits the scalability is the
over-partitioning of the mesh at higher machine scale. This leads to an increase in redundant computation at the halo
regions (compared to the non-halo elements per partition) and an increase in time spent during halo exchanges. Evidence
for this explanation can be gained by comparing the average number of halo elements and the average number of neighbors
per MPI process reported by OP2 after partitioning with PTScotch and RCB (see Table \ref{tab/halos}). We noted
these results from runs on HECToR, but the halo sizes and neighbors are only a function of the number of MPI processes
(where one MPI process is assigned to one partition) as the selected partitioner gives the same quality partitions for a
given mesh for the same number of MPI processes on any cluster. 

Column 4 and 7 of Table~\ref{tab/halos}(a) details the average total number of nodes and edges per MPI process
when partitioned with PTScotch and RCB respectively. Columns 5 and 8 (\%H) indicate the average proportion of halo nodes
and edges out of the total number of elements per MPI process while Columns 3 and 6 (Av) indicate the average number of
communication neighbors per MPI process. With PTScotch, the proportion of the halo elements out of the average total
number of elements held per MPI process ranged from about 5\% (at 32 MPI processes) to about 27\% (at 4096 MPI
processes). The average number of MPI neighbors per MPI process ranged from 7 to 10. The halo sizes with RCB  were
relatively large, starting at about 10\% at 32 MPI processes to about 40\% at 4096 processes. Additionally the average
number of neighbors per MPI process was also greater with RCB. These causes point to better scaling with PTScotch
which agrees with the results in Figure \ref{fig/scaling}(a). 

The above reasoning, however, goes contrary to the relative performance difference we see between OP2's MPI only and
MPI+OpenMP parallelizations. We expected MPI+OpenMP to perform better at larger machine scales as observed in
previous performance studies using the Airfoil CFD benchmark\cite{PARCO2013}. The reason was that larger partition
sizes per MPI process gained with MPI+OpenMP in turn resulting in smaller proportionate halo sizes. But for Hydra,
adding OpenMP multi-threading has caused a reduction in performance, where the gains from better halo sizes at
increasing scale have not manifested into an overall performance improvement. Thus, it appears that the performance
bottlenecks discussed in Section~\ref{subsubsec/openmp} for the single node system are prevalent even at higher machine
scales. To investigate whether this is indeed the case, Table \ref{tab/omp-blk_atscale} presents the number of
colors and blocks for the MPI+OpenMP runs at increasing scale on HECToR. 

\begin{table}[t]\small
\centering\vspace{-0pt}
\caption{Hydra strong scaling performance on HECToR, Number of blocks (nb)\\ and number of colors (nc) for MPI+OpenMP
and time spent in communications (comm) and computations (comp) for the hybrid and the pure\\ MPI implementation: 2.5M
edges, 20 iterations}
\vspace{0pt}
\begin{tabular}{l|rr|rr|rr}
Num of 	&
\multicolumn{2}{c|}{MPI+OMP}&\multicolumn{2}{c|}{MPI+OMP}&\multicolumn{2}{c}{MPI} \\
nodes		& nb	& nc	& comm 	& comp	& comm	& comp \\
		&	&	&(sec.) &(sec.) &(sec.) &(sec.)\\\hline
1		& 9980	& 17	& 1.33	& 16.6 	& 1.2 	& 13.2	\\\hline
2		& 4950	& 16	& 1.04	& 7.8	& 0.83	& 6.5	\\\hline
4		& 2520	& 17	& 0.57	& 3.3	& 0.36	& 3.14	\\\hline
8		& 1260	& 15	& 0.52	& 1.42	& 0.23	& 1.48	\\\hline
16		& 630	& 14	& 0.26	& 0.81	& 0.21	& 0.68	\\\hline
32		& 325	& 13	& 0.28	& 0.4 	& 0.13	& 0.38	\\\hline
64		& 165	& 10	& 0.32	& 0.21	& 0.12	& 0.2	\\\hline
128		& 86	& 12	& 0.52	& 0.11	& 0.15	& 0.12	\\\hline
\end{tabular}
\label{tab/omp-blk_atscale}\vspace{-0pt}
\end{table}\normalsize

The size of a block (i.e. a mini-partition) was tuned from 64 to 1024 for each run, but at higher scale (from upwards
of 16 nodes) the best runtimes were obtained by a block size of 64. As can be seen, the number of blocks (nb)
reduces by two orders of magnitude when scaling from 1 node up to 128 nodes. However within this scale the number of
colors remains between 10 to 20. These numbers provide evidence similar to the ones we observed on the Ruby
single node system in Section~\ref{subsubsec/openmp} where a reduced number of blocks per color results in poor load
balancing. The time spent computing and communicating during each run at increasing scale shows that although the
computation time reduces faster for the MPI+OpenMP version, its communication times increase significantly, compared
to the pure MPI implementation. Profiled runs of MPI+OpenMP indicate that the increase in communications time is in fact
due to time spent in \texttt{MPI\_Waitall} statements where due to poor load balancing, MPI processes get limited by
their slowest OpenMP thread. 

Getting back to Figure \ref{fig/scaling}(a), comparing the performance on HECToR to that on the GPU cluster JADE,
reveals that for the 2.5M edge mesh problem the CPU system gives better scalability than the GPU cluster. We believe
that similar to the Airfoil code's GPU cluster performance~\cite{InPar2012}, this comes down to GPU utilization issues:
the level of parallelism during execution. Since the GPU is more sensitive to these effects than the CPU (where the
former relies on increased throughput for speedups and the latter depends on reduced latency), the impact on performance
is more significant due to reduced utilization at increasing scale. Along with the reduction in problem size per
partition, the same fragmentation as we observed with the MPI+OpenMP implementation due to coloring is present.
Colors with only a few blocks have very low GPU utilization, leading to a disproportionately large execution time. This
is further complemented by the different number of colors on different partitions for the same loop, leading to faster
execution on some partitions and then the idling at implicit or explicit synchronization points waiting for the slower
ones to catch up. We further explore these issues and how they affect performance of different types of loops in Hydra 
later in Section~\ref{subsubsec/breakdownperf}.


\subsubsection{Weak Scaling}\label{subsubsec/weakscaling}

Weak scaling of a problem investigates the performance of the application at both increasing problem and machine size.
For Hydra, we generated a series of NASA rotor 37 meshes such that a near-constant mesh size per node (0.5M vertices) is maintained at
increasing machine scale. The results are detailed in Figure \ref{fig/scaling}(b). The largest mesh
size benchmarked across 16 nodes (512 cores) on HECToR consists of about 8 million vertices and 25 million edges in
total. Further scaling could not be attempted due to the unavailability of larger meshes at the time of writing.

With OPlus, there is about 8-13\% increase in the runtime of Hydra each time the problem size is doubled. With OP2, the
pure MPI version with PTScotch partitioning shows virtually no increase in runtime, while the RCB partitioning slows down
3-7\% every time the number of processes and problem size is doubled. One reason for this is the near constant halo
sizes resulting from PTScotch, but with RCB giving 7-10\% larger halos. The second reason is the increasing
cost of MPI communications at larger scale, especially for global reductions. Similar to strong scaling, the MPI-only
parallelization performs about 10-15\% better than the MPI+OpenMP version. The GPU cluster, JADE, gives the best
runtimes for weak scaling, with a 4-8\% loss of performance when doubling problem size and processes. It roughly
maintains a 2$\times$ speedup over the CPU implementations at increasing scale. Adjusting the experiment to compare one
HECToR node to one GPU (instead of a full JADE node with 2 GPUs) still shows a 10-15\% performance advantage for the
GPU.

\begin{figure*}[!t]
\begin{center}
\subfloat[Strong Scaling (2.5M edges)
\label{fig/s_scaling}]{\includegraphics[width=8.5cm]{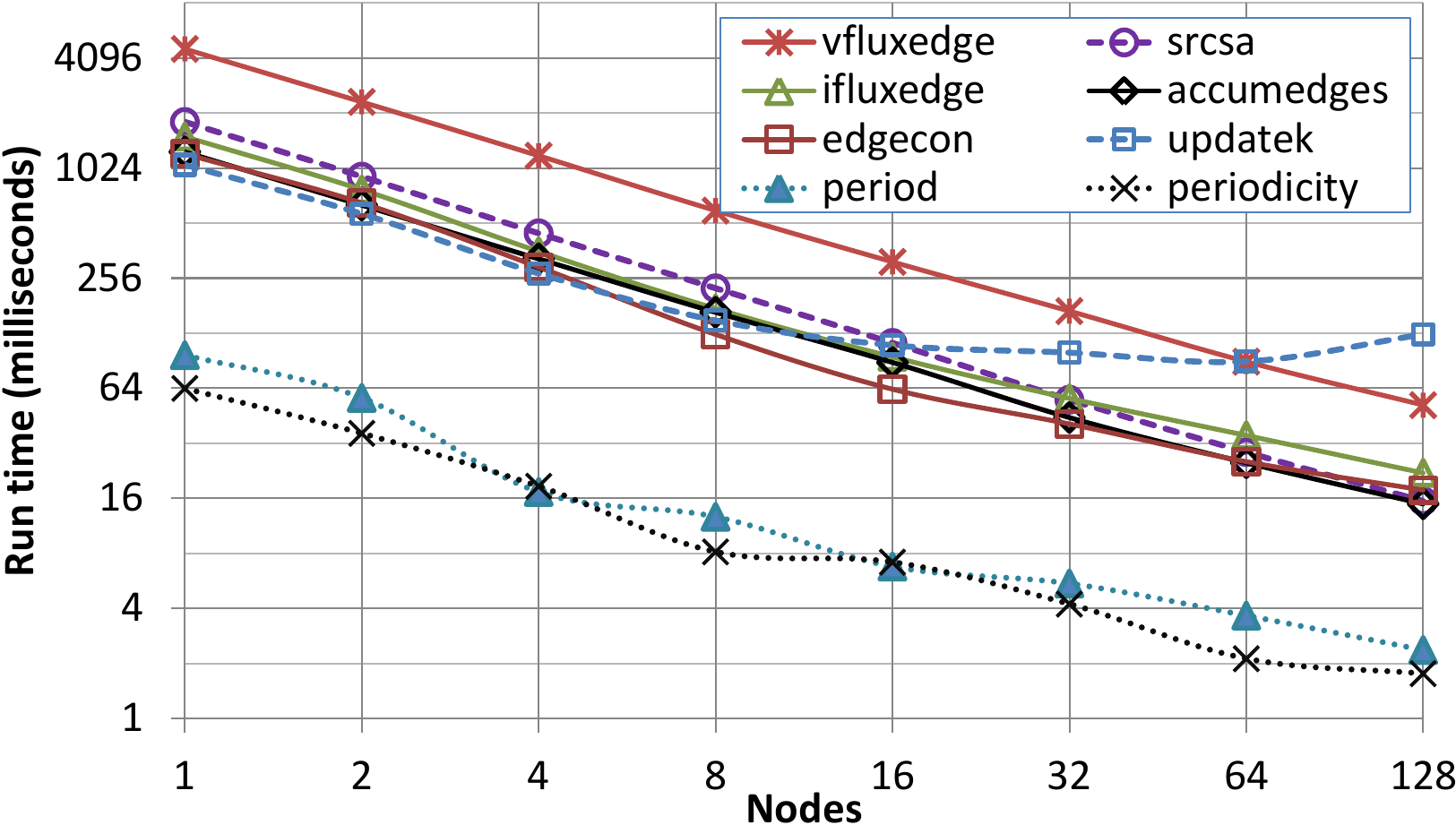}}\vspace{5pt}
\subfloat[Weak Scaling (0.5M edges per node)
\label{fig/w_scaling}]{\includegraphics[width=8.5cm]{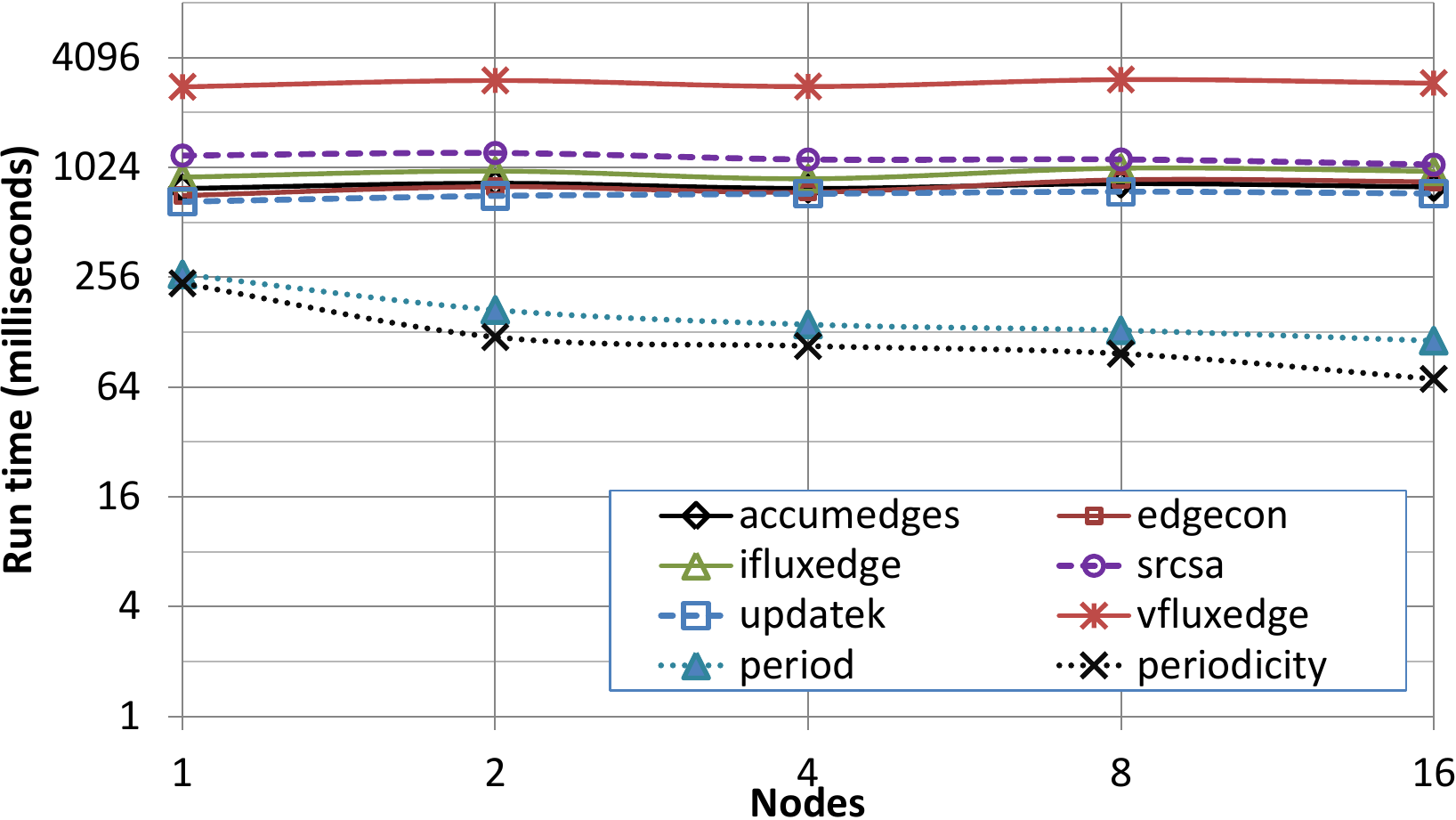}}\vspace{-5pt}
\caption{Scaling performance per loop runtime breakdowns\\ on HECToR (NASA Rotor 37, 20 iterations)}%
\end{center}\vspace{-15pt}
\end{figure*}

The above scaling results give us considerable confidence in OP2's ability to give good performance at large machine
sizes even for a complex industrial application such as Hydra. We see that the primary factor affecting performance is
the quality of the partitions: minimizing halo sizes and MPI communication neighbors. These results illustrate that,
in conjunction with utilizing state-of-the-art partitioners such as PTScotch, the halo sizes resulting from OP2's owner-compute design for distributed memory parallelization provide excellent scalability. We also see that GPU clusters are
much less scalable for small problem sizes and are best utilized in weak-scaling executions.


\subsubsection{Performance Breakdown at Scale}\label{subsubsec/breakdownperf} 

In this section we delve further into the performance of Hydra in order to identify limiting factors. The
aim is to break down the scaling performance to gain insights into how the most significant loops in the application
scale on each of the two cluster systems.

Figure \ref{fig/s_scaling} shows the timing breakdowns for a number of key loops when, for the MPI only version
(partitioned with PTScotch). Note how the loops \texttt{vfluxedge, edgecon} and \texttt{srcsa} at small scale account
for most of total runtime. However as they are loops over either interior vertices or edges, that do not include any
global reductions, they have near-optimal scaling. The loop \texttt{updatek} contains global reductions, and thus at
scale, it is bound by the latency of communications. At 128 nodes (4096 cores) it becomes the single most expensive
loop. Loops over boundary sets, such as \texttt{period} and \texttt{periodicity} scale relatively worse than loops over
interior sets, since fewer partitions carry out operations over elements in those sets.

\begin{figure*}[t]
\begin{center}
\subfloat[Strong Scaling (2.5M edges)\label{fig/s_scaling_gpu}]
{\includegraphics[width=8.5cm]{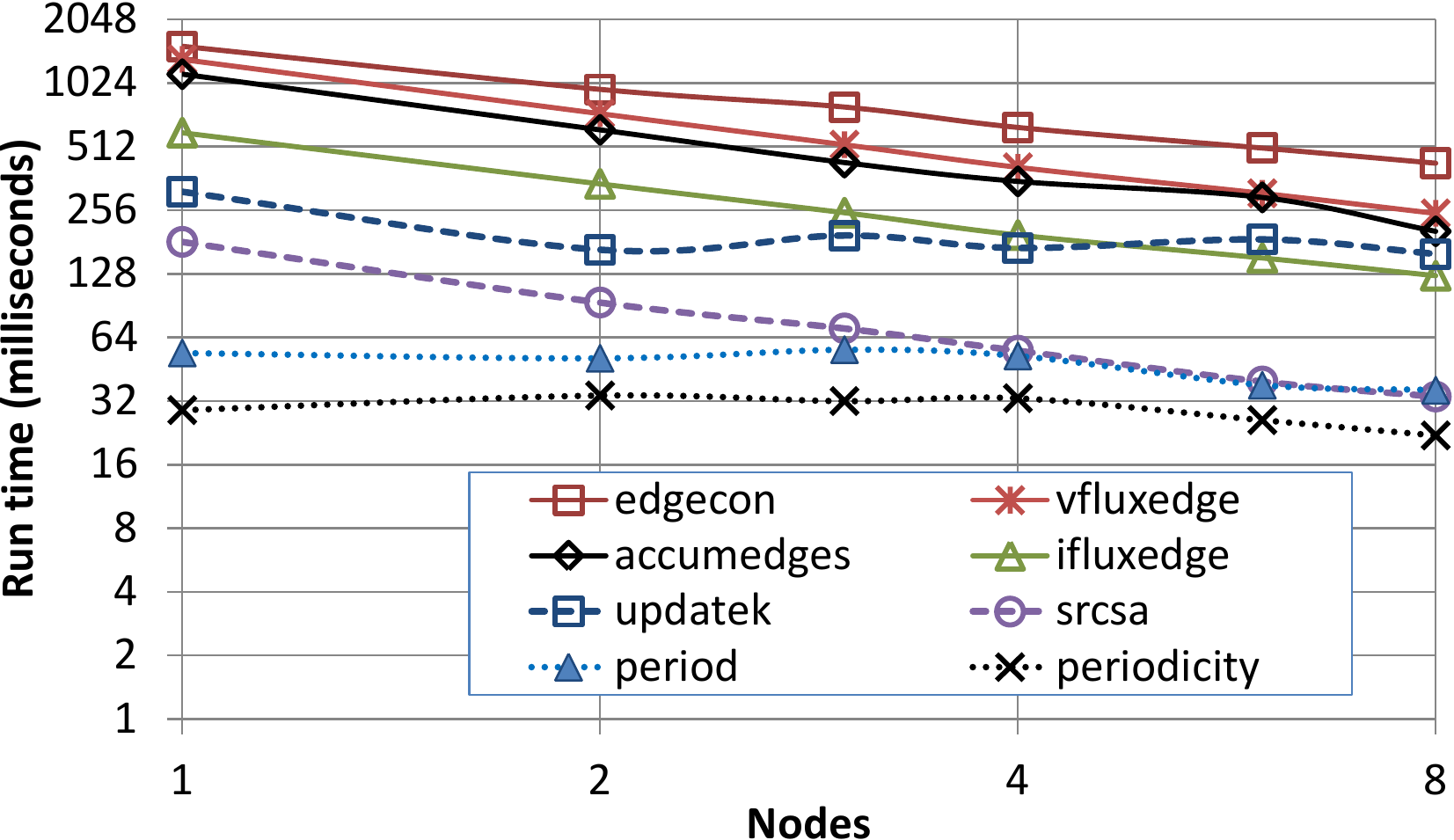}}\hspace{5pt}
\subfloat[Weak Scaling (0.5M edges per node)\label{fig/w_scaling_gpu}]
{\includegraphics[width=8.5cm]{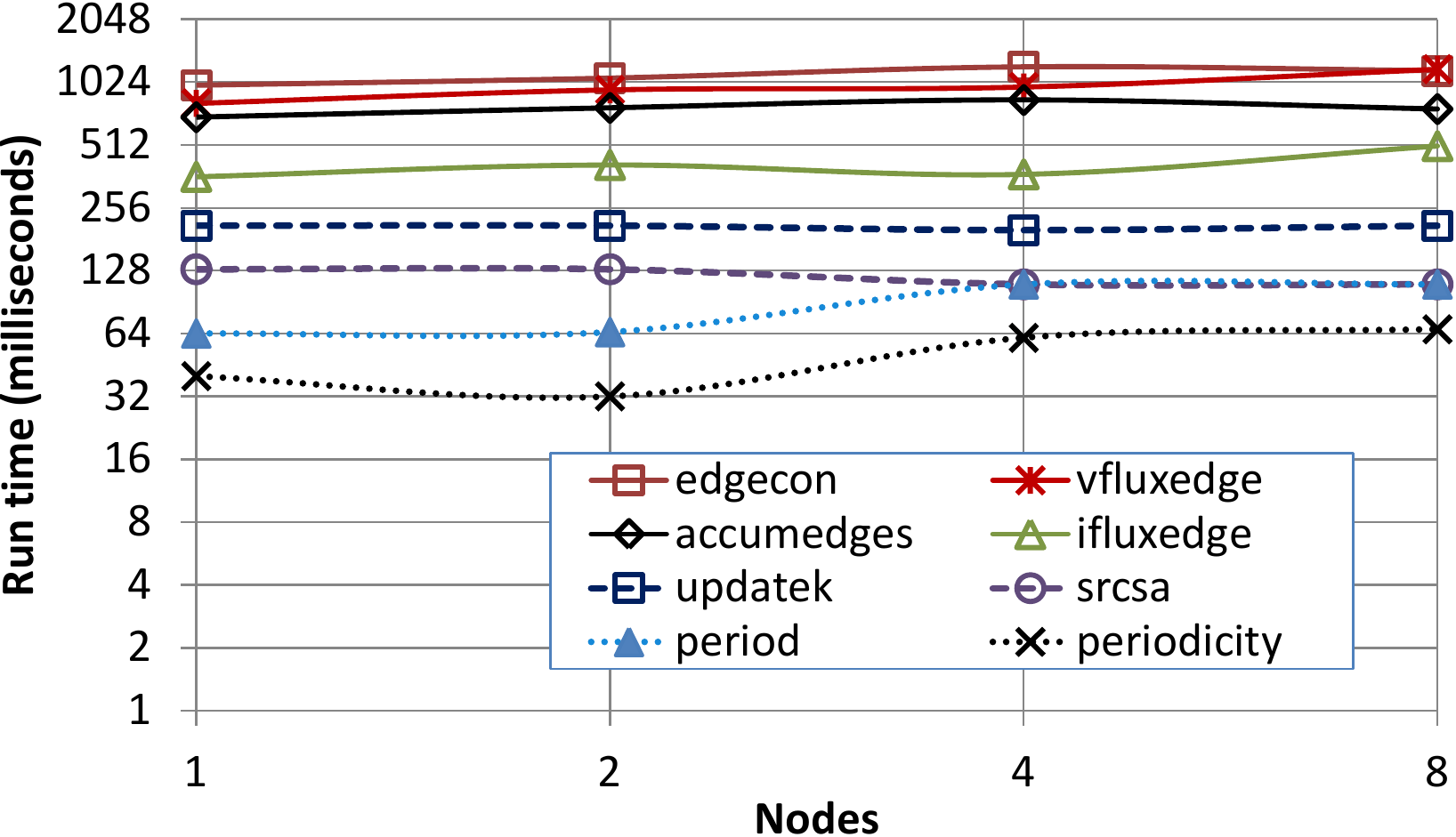}}\vspace{-5pt}
\caption{Scaling performance per loop runtime breakdowns\\ on JADE (NASA Rotor 37, 20 iterations)}%
\end{center}\vspace{-15pt}
\end{figure*}

Per-loop breakdowns when strong scaling on the Jade GPU cluster are shown in Figure \ref{fig/s_scaling_gpu}.
Observe how the performance of different loops is much more spread out compared to those on the CPU cluster scaling (as
shown in Figure \ref{fig/s_scaling}). Also note how boundary loops such as \texttt{period} and \texttt{periodicity} are
not so much faster than loops over interior sets, which is again due to GPU utilization. While the loop with reductions
(\texttt{updatek}) was showing good scaling on the CPU up to about 512 cores, performance stagnates beyond 4 GPUs which
is a result of the near-static overhead of on-device reductions and the transferring of data to the host, all of which
are primarily latency-limited. Most other loops, such as \texttt{vfluxedge, ifluxedge, accumedges} and \texttt{srcsa},
scale with 65-80\% performance gain when the number of GPUs are doubled, however \texttt{edgecon} only shows a 48-60\%
increase due to the loop being dominated by indirect updates of memory and an increasingly poor coloring at scale.


Figure \ref{fig/w_scaling} shows timing breakdowns for the same loops when weak scaling on HECToR, with very little
increase in time for loops over interior sets and a slight reduction in time for boundary loops, as a result of the
boundary (surface) of the problem becoming smaller relative to the interior. Similar results can be observed in
Figure \ref{fig/w_scaling_gpu} when weak scaling on the GPU cluster.  Here, some of the bigger loops gets
relatively slower, due to the load imbalance between different GPUs. In this case some partitions need more colors than
others for execution, which in turn slows down execution. Boundary loops such as \texttt{period} and
\texttt{periodicity} become slower as more partitions share elements of the boundary set forcing halo exchanges that are
limited by latency.

\section{Full Hybrid GPU-CPU execution}\label{sec/hybrid}

Most related work published on many-core acceleration, and GPU acceleration in particular, focuses on migrating the
entire code base to the GPU and then comparing performance with the CPU. However, modern GPU supercomputers, such as
Titan at Oak Ridge NL~\cite{titan}, consist of roughly the same number of GPUs and CPU sockets, and often pricing is
only calculated on a per-node basis. Thus, if an application only exploits the computational resources of the GPUs, then
the CPUs are idling, even though they might have considerable computational power themselves; this is a waste of energy
and resources. Several papers address this issue by employing different techniques, where the CPU and the GPU either
have the same ``rank'', such as in the case of shared task-queues~\cite{hybrid_taskqueue}, or where the GPU computes on
the bulk of the workload and the CPU handles the parts where the GPU would be underutilized, such as the boundary in
domain decomposition systems \cite{hybrid_atmospheric,hybrid_ms,hybrid_gromacs}. In this section we present preliminary
results on OP2's support for what we call full-hybrid execution where both the CPUs and the GPUs on a node is used for
mesh computations. Again, the possibility of seamlessly integrating such a fully-hybrid parallelization to Hydra is only
possible due to the high-level abstraction approach implemented through OP2.

The natural approach to enable hybrid CPU-GPU execution in OP2 is to assign some processes to execute on the GPU
and others to execute on the CPU. This hardware selection happens at runtime: on a node with $N$ GPUs, the first $N$
processes assigned to it pick up a GPU and the rest become CPU processes. To enable hybrid execution, the generated
kernel files include code for execution with both MPI+CUDA and MPI+OpenMP, thus at runtime the different MPI processes
assigned to different hardware can call the appropriate one.

The most important challenge with hybrid execution in general, not just for Hydra, is to appropriately load balance
between different hardware so that both are utilized as much as possible. From our results in the previous sections, the
importance of load balancing, even for executing computations the CPU only, was evident. Finding such a balance for
simple applications where one computational phase (such as a single loop) dominates the runtime may not be difficult.
One only needs to compare execution times on the CPU and the GPU separately and assign proportionally sized partitions
to the two. 

However, for an application such as Hydra consisting of several phases of computations, such a load balancing is
not trivial: the performance difference between the CPU and the GPU varies widely for different loops, as shown in
Table \ref{tab/loopperfOMP} and Table \ref{tab/loopperfCUDA}. For example the loop \texttt{vflux} is about
3 times faster on the GPU, but the loop \texttt{edgecon} is 25\% faster on the CPU. Load balancing for each
computational loop is infeasible as it would require repartitioning the mesh and transferring large amounts of data
between different processes from one loop to another, losing any performance benefit it might offer. Thus in OP2 we have
to come up with a static load balance upfront, which implies that some loops will be executed faster on the GPU than
the CPU and vice versa, leading to the faster one waiting for the slower one to catch up whenever a halo exchange is
necessary between them. This can severely restrict the potential performance gains expected from a fully
hybrid execution. 

\begin{figure}[t]\centering
\includegraphics[width=8.4cm]{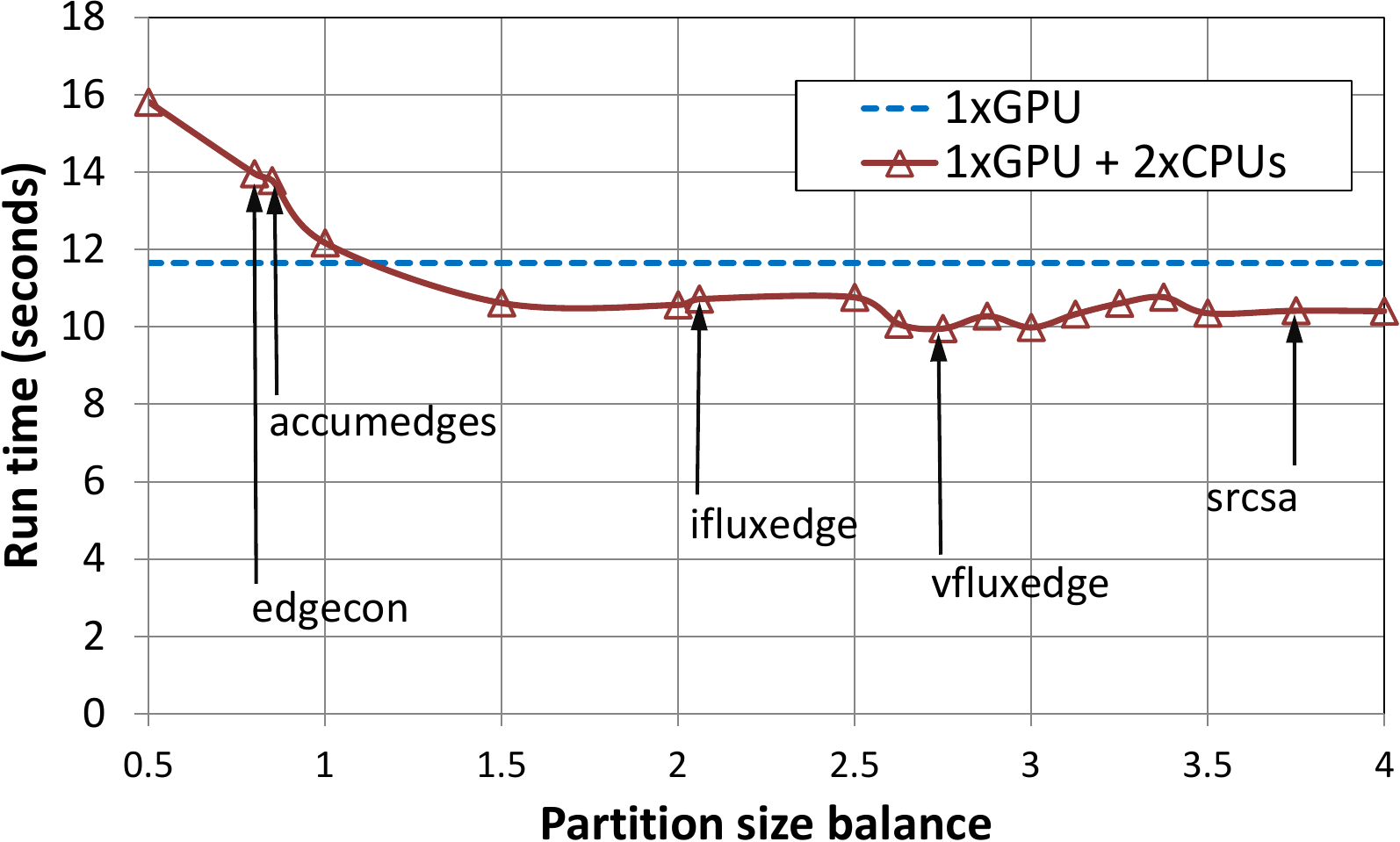}
\vspace{-5pt}\caption{Hybrid CPU-GPU performance at different load balancing values, marking perfect balance
for different loops on Ruby (NASA Rotor 37, 2.5M edges,\\ 20 iterations) with ParMetis partitioning}
\label{fig/Hybrid}\vspace{-15pt}
\end{figure}

We perform hybrid tests on the Ruby development machine, using a single GPU and both CPU sockets, each running OpenMP
with 10 threads. Partitioning is carried out using the heterogeneous load balancing feature of ParMetis.
Figure \ref{fig/Hybrid} shows performance while adjusting the static load balance between the work assigned to the
CPU and the GPU. For example a partition size balance of 2.5 implies that the GPU executes a partition that is 5
times larger than a single CPU (i.e. 2.5 times larger than the combined size of the partitions assigned to both of the CPUs
on Ruby).  The first guess for the static load balance would be based on single GPU execution time and MPI+OpenMP
execution time on Ruby, yielding a balance of about 1.5, giving a 9\% performance improvement over single GPU runtime.
We have carried out an auto-tuning run for the value of the static load balance, yielding the data points in
Figure \ref{fig/Hybrid}, registering the execution times of different loops on the CPU and the GPU. 

The figure marks load balance values where different loops are in perfect balance (i.e. the execution time on the GPU
and the two CPUs are approximately the same); it is obvious that there is a big variation, but the trends show that it
is best to shift the balance towards loops where the GPU has a significant speedup over the CPU, such
as \texttt{ifluxedge, vfluxedge} or \texttt{srcsa}. This increases the performance gain from hybrid execution up to 15\%.

\section{Related Work}\label{sec/relatedwork}

\noindent There are several well established conventional libraries supporting unstructured mesh based application
development on traditional distributed memory architectures. These include the popular PETSc~\cite{petsc},
Sierra~\cite{sierra} libraries as well as others such as Deal.II~\cite{deal2}, Dune~\cite{dune} and
FEATFLOW~\cite{featflow}. Recent research in this area include unstructured mesh frameworks that allows extreme scaling
(up to 300K cores) on distributed memory CPU clusters~\cite{CUBoulderAndRPI} while libraries such as PETSc have also
implemented hand tuned back-ends targeting solvers on distributed memory clusters of GPUs. Other related work
include RPI's FMDB~\cite{fmdb} and LibMesh~\cite{libMeshPaper}. A major goal of the LibMesh library is to provide support
for adaptive mesh refinement (AMR) computations. There are also conventional applications such as computational
fluid dynamics (CFD) solvers TAU~\cite{tau}, OpenFOAM~\cite{openfoam} or Fluent~\cite{fluent} that support
GPU acceleration of parts of the simulation, however these take the approach of having separate hand-written
implementations for different hardware. In contrast to these libraries, OP2's objective is to support multiple back-ends
(particularly aimed at emerging multi-core/many-core processor systems) for the solution of mesh-based applications
without the intervention of the application programmer.

OP2 can be viewed as an instantiation of the AEcute (access-execute descriptor)~\cite{AEcute} programming model that
separates the specification of a computational kernel with its parallel iteration space, from a declarative
specification of how each iteration accesses its data. The decoupled Access/Execute specification in turn creates the
opportunity to apply powerful optimizations targeting the underlying hardware. A number of research projects have
implemented similar or related programming frameworks. Liszt~\cite{liszt} and FEniCS~\cite{fenics} specifically target
mesh based computations.

The FEniCS~\cite{fenics} project defines a high-level language, UFL, for the specification of finite element algorithms.
The FEniCS abstraction allows the user to express the problem in terms of differential equations, leaving the details of
the implementation to a lower-library. Currently, a runtime code generation, compilation and execution framework that is based
on OP2, called PyOP2~\cite{pyop2}, is being developed at Imperial College London to enable UFL declarations to use
similar back-end code generation strategies to OP2. Thus, the performance results in this paper will be directly applicable to the performance of code
generated by FEniCS in the future. 

While OP2 uses an active library approach utilizing code transformation, Delite ~\cite{EPFL2011} and
Liszt~\cite{liszt} from Stanford University take the external approach to domain specific languages that require
advanced compiler technology, but offer more freedom in implementing the language. Liszt targets unstructured grid
applications; the aim, as with OP2, is to exploit information about the structure of data and the nature of the
algorithms in the code and to apply aggressive and platform specific optimizations. Performance results from a range of
systems (a single GPU, a multi-core CPU, and an MPI based cluster) executing a number of applications written using
Liszt have been presented in~\cite{liszt}. The Navier-Stokes application in~\cite{liszt} is most comparable to OP2's
Airfoil benchmark~\cite{PER2011} and shows similar speed-ups to those gained with OP2 in our work. Application
performance on heterogeneous clusters such as on clusters of GPUs is not considered in~\cite{liszt} and is noted as
future work.

There is a large body of work that develops higher-level frameworks for explicit stencil based applications (structured
mesh applications) including Paraiso~\cite{Paraiso}, Ypnos~\cite{Ypnos}, SBLOCK~\cite{SBLOCK}, SDSL ~\cite{SDSL},
the HIPAcc framework ~\cite{Membarth}, Pochoir ~\cite{Pochoir}, Patus ~\cite{Patus}, Mint ~\cite{Mint}, Kranc
~\cite{Kranc}, and others~\cite{cohen,Holewinski}. In a structured mesh, the mesh is regular and the connectivity is
implicit, where computations are performed based on a stencil that define a regular access pattern. Based on the
stencil, the required computation (or a kernel) is used to compute a new element value from the current element value
and neighboring elements. Ypnos~\cite{Ypnos} is a functional, declarative domain specific language, embedded in Haskell
and extends it for parallel structured grid programming. The language introduces a number of domain specific abstract
structures, such as \textit{grids} (representing the discrete space over which computations are carried out),
\textit{grid patterns} (stencils) etc. in to Haskell, allowing different back-end implementations, such as C with MPI or
CUDA. Similarly, Paraiso~\cite{Paraiso} is a domain-specific language embedded in Haskell, for the automated tuning of
explicit solvers of partial differential equations (PDEs) on GPUs, and multi-core CPUs. It uses algebraic concepts such
as tensors, hydrodynamic properties, interpolation methods and other building blocks in describing the PDE solving
algorithms. In contrast, most other projects, such as SBLOCK~\cite{SBLOCK}, SDSL ~\cite{SDSL}, HIPAcc ~\cite{Membarth}
and Pochoir ~\cite{Pochoir} express computations as kernels applied to elements of a set and use compiler technologies
and  automatic source code generation to accelerate execution.

We are not aware of research into the applicability of the above DSLs to large-scale or industrial applications.

Recently, directive based approaches have emerged as a much publicized solution to programming heterogeneous systems.
Most notable of these, OpenACC~\cite{openacc} attempts to follow the accessibility and success of OpenMP in programming
parallel systems but specifically targets accelerator based systems. Our experience suggests a range of extensive
language and compiler-technology improvements would be needed to enable applications with significant non-affine array
accesses (such as unstructured mesh applications with indirect array accesses) to provide the best performance in a
directive-based programming framework. The alternative, in which a significant effort is made by application developers
to explicitly optimize their code, is not a viable solution for an industrial code, due to its size and complexity. The
abstractions provided by the OP2 library (particularly the \texttt{op\_par\_loop} construct) can be seen as encompassing
not only the ``parallel regions'' of a directive based program that can be accelerated, but also an explicit description
of how each data item within the parallel region is accessed, modified and synchronized with respect to the unstructured
mesh data and computation it represents. With such an explicit \textit{access-descriptor}, OP2 allows for optimization
and parallel programming experts to choose significantly more  radical implementations for very specific hardware in
order to gain near-optimal performance.



\section{Conclusions}\label{sec/conclusions}

Rolls Royce's Hydra is a full-scale industrial CFD application currently in regular production use. Porting it to
exploit multi-core and many-core parallelism presents a major challenge in data organization and movement requiring the
utility of a range of low-level platform specific features. The research presented in this paper illustrates how the OP2
high-level domain specific abstraction framework can be used to future-proof this key application for continued
high performance on such emerging processor systems. We charted the conversion of Hydra from its original hand-tuned
production version to one that utilizes OP2, and mapped out the key difficulties encountered in the process. Over the course of
this process OP2 enabled the application of increasingly complex optimisations to the whole code to achieve near optimal
performance. The paper provides evidence of how OP2 significantly increases developer productivity in this task. 

Performance results for the code generated with OP2 demonstrate that not only could the same runtime performance as that of the
hand-tuned original production code be achieved, but it can be significantly improved on conventional processor
systems as well as further accelerated by exploiting many-core parallelism. We see that OP2's MPI and MPI+OpenMP
parallelizations achieve about 2 $\times$ speedup when strong-scaled and maintain 15-30\% speedup when weak-scaled
over the original implementation on a large distributed memory cluster system. Running on NVIDIA GPUs, OP2's CUDA
implementation achieves about 2 to 3 $\times$ speedups over the latest Intel Sandy Bridge x86-64 processors and
maintains a similar performance advantage over the CPU cluster implementations when weak-scaling over a GPU cluster. We
also demonstrate that executing Hydra on both the CPUs and GPUs in a fully-hybrid setting provides up to 15\% speedup
over the purely GPU execution and is primarily bound by load-balancing issues.

Some of the key optimisations that
affect all backend implementations are the use of improved partitioning methods, mesh renumbering for improved cache
locality and partial halo exchanges. Furthermore, for shared memory parallelism techniques, we have shown the importance
of optimizing the mini-partition size so as to have a balance of parallelism and data locality, and we presented further
techniques involving data layout transformations and the fine-tuning of resource usage to improve performance on the
GPU. 

We develop a deeper understanding of execution efficiency by investigating performance characteristics of the major
computational loops in Hydra. We show how near-optimal performance is achieved in most loops and identify bottlenecks
due to MPI collectives or inefficiencies in the execution scheme that are going to be a subject of future research.

We believe that the future of numerical simulation software development is in the specification of algorithms
translated to low-level code by a framework such as OP2. Such an approach will, we believe, offer revolutionary
potential in delivering performance portability and increased developer productivity. This, we predict, will be an
essential paradigm shift for addressing the ever-increasing complexity of novel hardware/software technologies. 

The full OP2 source and example benchmark applications are available as open source software~\cite{op2,op2-github} and
the developers would welcome new participants in the OP2 project.


\begin{acks}
This research has been funded by the UK Technology Strategy Board and Rolls-Royce plc.~through the Siloet project, the UK
Engineering and Physical Sciences Research Council projects EP/I006079/1, EP/I00677X/1 on ``Multi-layered Abstractions
for PDEs'' and the ``Algorithms, Software for Emerging Architectures`` (ASEArch) EP/J010553/1 project. The authors would
like to acknowledge the use of the University of Oxford Advanced Research Computing (ARC) facility in carrying out this
work. 

The authors would like to thank PGI and Brent Leback for their proactive support in improving their CUDA Fortran
compiler and Maxim Milakov of NVIDIA for his help and advice on GPU optimizations. 

We are thankful to Leigh Lapworth, Paolo Adami and Yoon Ho at Rolls-Royce plc. for providing access to the Hydra CFD
application, Gr aham Markall, Fabio Luporini, David Ham and Florian  Rathgeber, at Imperial College London, Lawrence
Mitchell at theUniversity of Edinburgh for their contributions to the OP2 project and Endre L\'{a}szl\'{o}, Andr\'{a}s
Ol\'{a}h and Zolt\'{a}n Nagy at PPKE Hungary.
\end{acks}

\bibliographystyle{ACM-Reference-Format-Journals}
\bibliography{hydra.bib}


\end{document}